\documentclass[twocolumn,superscriptaddress,
 amsmath,amssymb,
 aps,
 prl,
]{revtex4}
\usepackage{dcolumn}
\usepackage{bm}
\usepackage{graphicx}
\usepackage{comment}
\usepackage[caption=false]{subfig}
\usepackage{color}

\usepackage[colorlinks=true,citecolor=blue,urlcolor=red]{hyperref}
\usepackage{physics}
\usepackage{siunitx}
\usepackage{cancel}
\usepackage{ulem}

\newcommand{\ser}[1]{\textcolor{black}{#1}}

\begin{document}

\title{Experimental evidence for strong emergent correlations between particles in a switching trap}

\author{Marco Biroli}
\affiliation{LPTMS, CNRS, Univ.  Paris-Sud,  Universit\'e Paris-Saclay,  91405 Orsay,  France}
\author{Sergio Ciliberto}
\affiliation{Laboratoire de Physique, CNRS, ENS de Lyon, F-69342 Lyon, France}
\author{Manas Kulkarni}
\affiliation{ICTS, Tata Institute of Fundamental Research, 560089 Bengaluru, India}
\author{Satya N. Majumdar}
\affiliation{LPTMS, CNRS, Univ.  Paris-Sud,  Universit\'e Paris-Saclay,  91405 Orsay,  France}
\author{Artyom Petrosyan}
\affiliation{Laboratoire de Physique, CNRS, ENS de Lyon, F-69342 Lyon, France}
\author{Gr\'egory Schehr}
\affiliation{Laboratoire de Physique Th\'eorique et Hautes Energies, CNRS UMR 7589, Sorbonne Universit\'e, 4 Place Jussieu, 75252 Paris Cedex 05, France}

\date{\today}

\begin{abstract}
We experimentally study a \ser{system of $N = 4$ two-dimensional Brownian particles}, each confined in a harmonic trap with identical stiffness. The stiffness switches \textit{simultaneously} between two values at random Poissonian times. This collective switching drives the system into a non-equilibrium stationary state (NESS) with strong long-range correlations between the positions of the particles. \ser{Remarkably, we find that, despite the presence of hydrodynamic interactions between the particles mediated by the surrounding fluid, the statistics of some observables are insensitive to hydrodynamic interactions and are well described by the noninteracting theory.}  
Comparing with exact theoretical predictions for noninteracting particles, we observe excellent agreement between theory and experiments for three such observables, namely the correlations between particles, extreme value and order statistics (maxima, minima and ranked positions) and the full counting statistics (i.e., the distribution of the number of particles in a finite interval $[-L, L]$ around the trap center).
\end{abstract}

\maketitle


Most of physical systems, such as particles, spins, etc, are usually coupled to an external environment, which can be just a hard box or a heat bath. 
In typical problems, one assumes that the environment is ``big'' and does not fluctuate with time. It is thus characterised by some fixed parameters, such as the size of the box or the temperature. However, in many situations, the environment may not necessarily be big and its stochastic fluctuations may induce strong correlations between the particles, even though the particles may not have any direct interactions between them. 
A classical example goes back to { Huygens'} pendulum in the seventeenth century where he conducted a simple experiment with two pendulums hung from a common wooden beam, sitting on two opposite chairs (for a historical account see \cite{huygens_book}). The two pendulums do not have any direct interaction between them, but when one perturbs only one of the pendulums, then Huygens observed that, after some time, both pendulums start oscillating synchronously. In this simple example, the perturbed pendulum affects the environment, i.e., the wooden beam in this example, which in turn imparts a motion to the other pendulum. In this example, the
two pendulums get strongly correlated through the dynamics of the environment. 

There are however situations where the particle motions do not affect the environment, but nevertheless the stochastic fluctuations of the environment can induce strong {dynamically emergent} correlations between noninteracting particles, simply due to the fact that the particles share the same fluctuating environment. There have been few theoretical studies, mostly in one-dimension, on this problem. This includes noninteracting particles undergoing zero temperature gradient descent dynamics on a $1+1$ dimensional fluctuating interface belonging to the Edwards-Wilkinson or the Kardar-Parisi-Zhang universality class \cite{NBM05,NBM06}. Another well studied system corresponds to particles in a trap whose shape changes stochastically with time. A single particle driven by such a stochastically switching trap reaches 
a non-equilibrium stationary state (NESS) at long times. This has been demonstrated in several one-dimensional models~\cite{MBMS20,GK21,SDN21,MBM22}. 

When there is more than one particle in such a switching trap, the particles may get correlated in their NESS even though there may not be any direct interactions between the particles. This was demonstrated recently in an exact solution of a model 
of $N$ noninteracting Brownian particles in a harmonic trap whose stiffness switches between two values with some rates~\cite{BKMS24}. There, it was found that the system reaches a NESS at long times, where the particles get strongly correlated, due to the stochastic switching of the stiffness of the harmonic trap. Other theoretical models, both classical and quantum, have recently been studied where the stochastic fluctuations of the environment were found to induce strong {dynamically emergent} correlations that persisted all the way up to the NESS~\cite{BLMS23,BLMS24,SM2024,KMS2025,MMS2025}. The purpose of this paper is to study experimentally such dynamically growing correlations between particles emerging from the stochastic fluctuations of the shared common environment.

\begin{figure}
    \centering
    \includegraphics[width=0.4\textwidth]{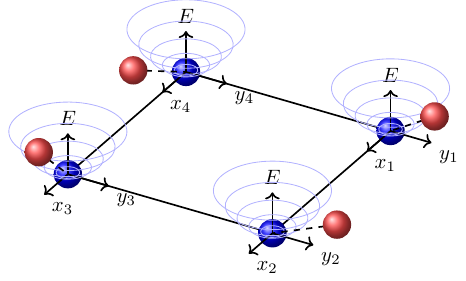}
    \caption{A sketch of the experimental setup considered in this Letter. Four colloidal particles (of radius $1\mu$m) are placed at the vertices (marked blue) of a square of size $\sim 6\mu$m. Using a fast sweeping laser we create four harmonic traps confining the particles close to their respective vertex. At random Poissonian times we switch the stiffness of the trap by modulating {the laser intensity, at the objective  entrance, from $4$ to $40$mW }. Using a camera we track the position of each particle dynamically (with the current positions represented schematically by red spheres) in the $(x_i, y_i)$ plane.  Here, $E$ represents the energy axis of the parabolic potential.}
    \label{fig:enter-label}
\end{figure}


For the purpose of the experiment, we focus on the theoretical example of $N$ noninteracting particles in a switching harmonic trap in one-dimension~\cite{BKMS24}. The stiffness of the trap switches between values $k_1$ and $k_2$ with rates $r_1$ (from $k_1$ to $k_2$) and $r_2$ (from $k_2$ to $k_1$). Here the stochastic switching plays the role of the environmental fluctuations that drive the system to a NESS at long times. 
The joint distribution of the positions of the particles in the NESS, despite being strongly correlated, displays an interesting solvable structure that enables analytical predictions for a number of physically measurable observables, such as the average density profile, the extreme value statistics, the spacing statistics and the full counting statistics, etc.~\cite{BKMS24}. However, these analytical predictions~\cite{BKMS24} crucially assume that the particles are non-interacting. In real systems, such as colloids in a liquid, there are typically long-range interactions.   
Examples include hydrodynamics~\cite{Berut,kotar,Barlett,polin}, magnetic~\cite{gao}, electric~\cite{dobnikar}, Janus colloidal particles~\cite{yan} and critical Casimir forces~\cite{ignacio}. In our system of colloidal particles in water, these long-range interactions are hydrodynamic in nature (see for example~\cite{Berut}). 
However, such long-range interactions between the particles were not taken into account in the theoretical computation in Ref.~\cite{BKMS24}. A natural question then is how the direct hydrodynamic interactions between the particles may modify the NESS.  

One particularly interesting limit of this model corresponds to taking $k_1 \to +\infty$, $k_2 \to 0$, $r_1 \to +\infty$, and $r_2 \to r$. In this limit, the model corresponds to $N$ independent Brownian motions undergoing a simultaneous stochastic resetting to the origin with rate $r$~\cite{BLMS23,BLMS24}. Stochastic resetting of a single diffusing particle has been studied quite extensively in the statistical physics literature in the recent past, both theoretically~\cite{EM11a,EM11b,EMS20,PKR22,GJ22}  and experimentally~\cite{Roichman20,BBPMC_20,FBPCM_21,Landauer23}. Multiple particles with and without interactions, subjected to stochastic resetting, have also been studied theoretically in a variety of systems~\cite{DHP14,GMS14,BBR16,MSS18,RTLG18,BKP19,MMS20,PCPL21,NG23,BMS23,BMS25}. 
In Refs.~\cite{BLMS23,BLMS24} the joint distribution of the positions of the Brownian particles and their correlations in the NESS, as well as the other physical observables discussed above, were computed analytically. In this example, the correlations emerge from the simultaneous resetting. In contrast, if the particles are reset independently, they remain independent at all times. Very recently, this limiting case was probed 
experimentally in a system of colloidal particles which were mechanically reset to the origin, using optical tweezers~\cite{VR25}. Interestingly, it was found that the hydrodynamic interactions have a less pronounced effect in the case of simultaneous resetting, compared to the case of independent resetting~\cite{VR25}. It is then natural to investigate whether the same conclusion holds in the more physically relevant and general  setting where $k_1, k_2, r_1$ and $r_2$ are all finite. 

Detecting such dynamically emergent correlations (DEC) in real experimental systems however poses a significant challenge, as they may be completely masked by the existent {long-range} hydrodynamic interactions between the particles. A priori, there are three possible scenarios:
(i) the hydrodynamic interactions govern the stationary-state properties, fully masking the DEC and rendering them experimentally inaccessible;
(ii) the dynamically generated correlations determine the stationary behavior, allowing the theoretical predictions of Ref.~\cite{BKMS24}, which neglect direct interactions, to accurately describe the system;
(iii) both hydrodynamic interactions and DEC are relevant in the steady state, and neither can be neglected.
Surprisingly, our experiments reveal that scenario (ii) holds: the stationary properties of several physical observables are well described by the non-interacting theory of Ref.~\cite{BKMS24}, despite the presence of hydrodynamic interactions.

Our experimental system, {as described in the Supplemental Material (SM) \cite{SM} and depicted in Fig.~\ref{fig:enter-label},} consists  of up to $N=4$ diffusing particles (silica beads in water with { a friction coefficient} $\nu = 1.88 \cdot 10^{-8} \; {\rm N \, s / m}$)  
in the presence of a  harmonic trap with intermittent stiffnesses switching between two adimensional values $k_1 > k_2$, measured in units of 
$k_0 = 10^{-6} {\rm N/m}$. The rates of switching $r_1$ and $r_2$ are also adimensional, measured in units of $1/\tau_0 = k_0 / \nu = 0.019  {\rm s}^{-1}$. 
In the experiment, we always set $r_1 = r_2 = r$, for simplicity. Each particle diffuses with a diffusion constant $D = k_B T / \nu$ (where $k_B$ is the Boltzmann constant) which is obtained experimentally from a reference variance $\sigma_0^2 = k_B T / k_0 = 4.07 \cdot 10^{-15} \, {\rm m^2}$ and the reference time $\tau_0$ as $D = \sigma_0^2 / \tau_0$. {In what follows, for the comparison between theory and experiment, we use $\sigma_0$ and $\tau_0$ as the units of length and time respectively. }

Before we describe our experimental setup, let us briefly recapitulate the theoretical predictions~\cite{BKMS24} for non-interacting particles in a switching harmonic trap as described above ({see \cite{SM} for a detailed discussion of the {switching} protocol}). The joint probability distribution function (JPDF) of the positions of the particles in the NESS can be written explicitly {\it for any} $N \geq 1$ as
%
$\tilde{P}(\{x_i\}) = P\big( \{z_i\} = \{x_i / \sigma_0\}\big) \sigma_0^{-N}$,
where the JPDF $P(\{z_i\})$ of the adimensional particle positions $\{z_i\}$ {(measured in units of $\sigma_0$)} is given by~\cite{BKMS24}
\begin{equation} \label{eq:jpdf}
    P(\{z_i\}) = \int_0^{1} \dd u \; h(u) \prod_{i = 1}^N \frac{e^{-\frac{z_i^2}{2 V(u)}}}{\sqrt{2 \pi V(u)}} \;,
\end{equation}
where $V(u) = \left[\frac{u}{k_2} + \frac{1 - u}{k_1}\right]$ and 
\begin{equation} \label{eq:def-h}
    h(u) = A \, u^{R_1 - 1} (1-u)^{R_2 - 1} V(u) \;,
\end{equation}
where $R_i = \frac{r_i}{2 k_i}$ with $i=1,2$. The function $h(u)$ is normalized to unity: $\int_0^{1}\dd u \;h(u) = 1$. The overall normalization constant $A$ can be explicitly computed and is given in the Supplementary Material (SM) \cite{SM}. The JPDF in Eq.~(\ref{eq:jpdf}) clearly does not factorise, indicating the presence of correlations between particles. The variance ${\rm Var}(z_i) = \langle z_i^2 \rangle$ is independent of $i$ and is also given in the SM. It reads 
\begin{equation} \label{Var}
{\rm Var}(z_i) = \langle z_i^2 \rangle = \frac{(4R_1R_2+R_1+R_2)}{r(1+R_1+R_2)} \;,
\end{equation} 
where $R_1 = r/(2 k_1)$ and $R_2 = r/(2 k_2)$. To probe the correlations between the positions of
two different particles labelled $i$ and $j$ (with $i \neq j$), we compute the simplest nonzero adimensional correlator 
\begin{equation} \label{C2}
C_{2}={\langle z_i^2 z_j^2 \rangle \over \langle z_i^2 \rangle \langle z_j^2\rangle}-1 \;.
\end{equation}
In the NESS, its exact value is given by (see SM \cite{SM})
\begin{eqnarray}	
C_2={(k_2-k_1)^2 (2+3 R_1+3 R_2+4R_1 R_2)\over (2+R_1+R_2)(2r+k_1+k_2)^2} \;.
	\label{eq_C2_norm_letter}
\end{eqnarray}
Note $R_1 = r/(2 k_1)$ and $R_2 = r/(2 k_2)$ also depend on $k_1$ and $k_2$ for fixed $r$. We would like to remark that the standard correlator $C_1=\langle x_ix_j\rangle/ \sqrt{\langle x_i^2\rangle \langle x_j^2\rangle}$ vanishes due to symmetry~\cite{BKMS24}, hence to detect the correlation, we need to compute a higher order correlator such as $\langle x_i^2x_j^2\rangle-\langle x_i^2\rangle\langle x_j^2\rangle$, or its normalized version $C_2$. {A comprehensive discussion of the roles of hydrodynamic interactions and switching on $C_1$ can be found in the End Matter and in the SM \cite{SM}.} The JPDF in Eq.~(\ref{eq:jpdf}) has a special conditionally independent and identically distributed (CIID) structure where the conditioning variable $u$ can be interpreted as the effective fraction of time the particles spend in the phase where the stiffness is $k_2$~\cite{BKMS24}. For a fixed $u$, the integrand has a product structure, making the particles independent. Thus, one can compute any observable of this correlated gas by first computing it for $N$ independent Gaussians with variance $V(u)$ and then average over $u$ drawn from $h(u)$. One important observable is the so called extreme value statistics and the order statistics~\cite{MS24}. This means probing the statistics of the position of the $K$-th rightmost particle in the gas in its NESS. For example, the distribution of the adimensional maximum, i.e., the {scaled} position of the rightmost particle {$\tilde{M}_1 =\max \{z_i\}$} can be computed from Eq. (\ref{eq:jpdf}) and is given in Eq. (\ref{eq:max}).
The corresponding result for the distribution $P(\tilde{M}_K,N)$ of the $K$-th scaled maximum $\tilde M_K = M_K/\sigma_0$ is given in Eqs. (S7)-(S8) of the SM \cite{SM}. Similarly, other observables such as the full counting statistics, i.e., the distribution $P(n_L,N)$ of the number of particles $n_L$ in a fixed interval $[-L, +L]$ around the trap center, can be computed exactly for any finite $N$~\cite{BKMS24}. The exact expressions are provided in Eqs.~(S9)-(S10) in the SM \cite{SM}.

\vspace{0.2cm}

\paragraph{Experimental setup. --} {We first note that, ideally, one would like to perform the experiment with all the particles in a {\it single} switching harmonic trap. However, this is not possible because all of the particles align on the laser beam optical axis  leading to extremely unstable situation induced by the particle repulsion  and by radiation pressure which pushes the particles out of the trap center. To circumvent this problem, we consider $N$ traps, each containing one particle and the stiffnesses of all the traps are modulated synchronously. The position of each particle is measured from the center of its own trap. If the particles were noninteracting, this would correctly mimic the system of $N$ independent particles in a single switching trap. In our system, we consider four such traps, each containing one silica bead     
(with radius $a = \SI{1}{\micro\meter} \pm 5\%$), trapped in water on the corners of a square of side  $l = \SI{6}{\micro\meter}$ by a laser beam (wavelength \SI{532}{\nano\meter}) which is periodically  moved on each of the four corners by two Acousto-Optic Deflectors (AOD) -- see Fig.~\ref{fig:enter-label}. {Note that,  although the trap centers are $\SI{6}{\micro\meter}$ apart, the fluctuations of each bead brings them to almost collide several times during the experiment.} The total  round trip time of the laser to visit the  four points is about \SI{1}{\milli\second} for the full round trip. The moving laser beam  is focused by  an oil-immersion objective (HCX PL. APO $63\times$/$0.6$-$1.4$) inside 
{a cell filled with a water-bead solution. The cell has dimensions $10\times$\SI{10}{\milli\meter}$^2$ and thickness \SI{250}{\micro\meter}. It is connected to a reservoir which contains the largest number of   beads  dispersed (during sample preparation) in bidistilled water at low concentration. }

{The reservoir and the low concentration 
do not allow other beads to be attracted into one of the four confining potentials of our system. If the other beads could get closer, that would result in perturbing the motion of the already trapped beads.  The probability of such accidental attractions is further reduced by moving the four trapped beads at  least \SI{5}{\milli\meter} away from the beads reservoir. This experimental arrangement}
allows us to perform  very long measurements of several hours without any perturbation from the other beads. The stiffness $k$ of the traps  can be changed from $0.2$ to $2\mu$N/m by changing  the laser intensity{, at the objective entrance} from \SI{4}{} to \SI{40}{\milli\watt}. The laser intensity  is controlled by the  amplitudes of the AOD driving voltages which determine the fraction of the light transferred into the deflected beam.  The beads are trapped at about $20\mu$m above the bottom plate of the cell in order to reduce the viscous interactions with the wall.
Thus in the absence of the other beads the relaxation time of a bead in the trap is simply \cite{Berut} $\tau=\nu/k$ where $\nu=6\pi \eta a$ and $\eta$ is the dynamic viscosity of water. The typical value  of  $\tau$ at $k=k_0=1\mu$N/m is $\tau_0\simeq0.02$s. 

The position of the beads is tracked by a fast camera with a resolution of \SI{115}{\nano\meter} per pixel which, after treatment, gives the position with an accuracy better than \SI{2}{\nano\meter}. \ser{We have $N=4$ two-dimensional particles. The $x$-coordinates $\{x_1, x_2, x_3, x_4\}$ provide us $N=4$ one-dimensional particles. Similarly the $y$-coordinates $\{y_1, y_2, y_3, y_4\}$ gives us another set of $N=4$ one-dimensional particles. For better statistics, we can think of each one representing a separate one-dimensional Brownian motion and, hence, we can effectively access $N_{\rm eff} = 8$ particles.} The four $(x_i,y_i)$ trajectories of the beads are sampled at \SI{1000}{\hertz}.   
The value of the stiffness for each trap $k_i$ is estimated from equipartition by measuring the variances, i.e., $k_i=k_BT/\langle x_i^2 \rangle$ where $k_{B}$ is the Boltzmann constant and $T$ is the temperature. {This estimation is correct even in the presence of the  other beads because the main interactions between the beads is viscous \cite{Berut} and  for the distances considered here the Coulombian interaction between the particle surfaces is negligible.} The dispersion $\Delta k_i$ of $k_i$ among the $8$ available directions  is mainly due to the dispersion of the bead size. Indeed we found that $\Delta k_i/k_i$ is at most $5\%$ but it can be  highly reduced to about $1\%$ for  good sets of beads. {In any case for the comparison with theory we used for $k$ the mean of the $8$ measured values of $k_i$}. 
The standard deviation around $k=k_0$ is $\sigma_0=\sqrt{k_BT/k_0}\simeq \SI{63}{\nano\meter}$.

 In our theoretical setup the trapping potentials of all the beads are simultaneously changed between two { stiffnesses} at a switching rates $r_1,r_2$.
 In the experiment we fix $r=r_1=r_2$ and we numerically generate {two signals} of amplitudes $A_1,A_2$ in which the residence time $t_r$ in $A_1$ and $A_2$ is exponentially distributed, i.e. $P(t_r)=r_e\exp(-r_e\ t_r)$ where $r_e=r/\tau_0$ is the dimensional value of $r$. This  numerically generated noise  is sampled at \SI{1000}{\hertz} and converted by a NI PXIe-6366 card to a voltage which is used to drive the AOD. As a consequence the laser intensity is modulated and the {stiffnesses} of the four traps change simultaneously between $k_1$ and~$k_2$. {We refer the reader to \cite{SM} for more details on the experimental setup.}  
 
 \paragraph{Data analysis.}  As a first check, we measured the standard deviation $\sqrt{{\rm Var}(z_i)}$ in the NESS given in Eq.~(\ref{Var}) as a function of $k_2$ (for fixed $k_1$ and $r$). The results are plotted in Fig. \ref{fig:std_C2_new} a) for two different values of $r$. The experimental results show a very good agreement with the theoretical formula in Eq.~(\ref{Var}). The larger error in the blue point (for $r=1.88$)  is due to the fact that the dispersion of the differences of the stiffness were larger than in the other measurements performed  at  $r=0.47$.
 
 Next, we consider our principal observable of interest, namely the nonzero correlator in the NESS given in Eq.~(\ref{eq_C2_norm_letter}). This correlation, as explained before, emerges from the dynamic switching of the trap stiffnesses between $k_1$ and $k_2$, with $k_1 \neq k_2$. Indeed, when $k_1 = k_2$, the trap has only one stiffness and in this case, at long times, the system reaches thermal equilibrium and the particles, in the absence of direct interaction between them, remains uncorrelated in the equilibrium stationary state. Consequently, for $k_1 = k_2$, one would expect that the correlator $C_2$ would vanish in equilibrium if there is no interaction between particles. Indeed, one sees from Eq.~(\ref{eq_C2_norm_letter}) that $C_2 = 0$ when $k_1 = k_2$. We have measured this correlator $C_2$ in equilibrium (in the absence of switching) and found indeed that it is very small, of the order $10^{-3}$. Then, we switch on the modulation of the traps between two values $k_1$ and $k_2$ and wait long enough time for the system to reach the stationary state (NESS this time). In the NESS we measure $C_2$ and compare it with the theoretical prediction given in Eq.~(\ref{eq_C2_norm_letter}). In Fig. \ref{fig:std_C2_new} b), we plot $C_2$ as a function of $k_2$ for fixed $k_1 = 1.44$ and for two different values of $r = 0.47$ and $r=1.88$. The measure confirms that when $k_2\rightarrow k_1$ then $C_2\rightarrow 0$. For all parameter values, we find an excellent agreement between theory and experiment. 
\begin{figure} 
	\centering
 \includegraphics[width=0.99\linewidth]{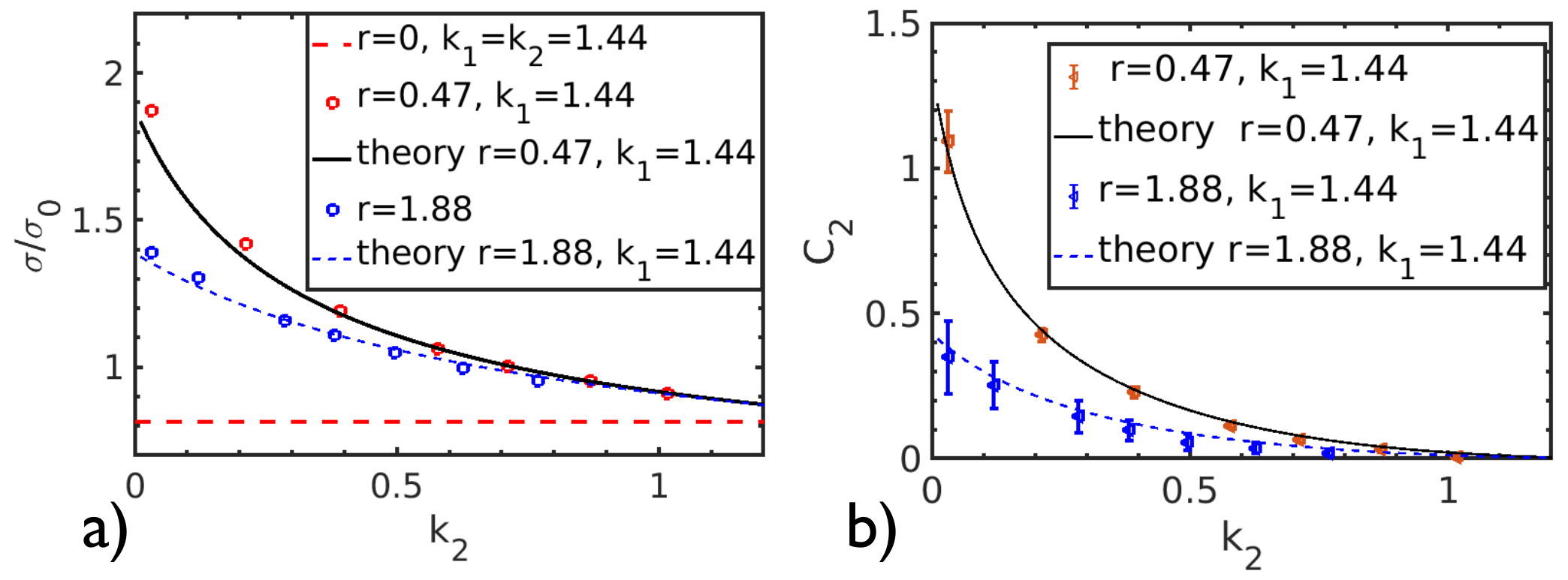}
	\caption{{\bf a)} Standard deviation  $\sqrt{{\rm Var}(z_i)}$ given in Eq.~(\ref{Var}) plotted as a function of $k_2$ for fixed $k_1=1.44$ and for two different values of $r=0.47$ (red circles) and $r=1.88$ (blue circles). The dashed lines represent the theoretical formula given in Eq.~(\ref{Var}), while the symbols represent the experimental data. {\bf b)} Correlation function $C_2$ defined in Eq.~(\ref{C2}), plotted as a function of $k_2$ for fixed $k_1=1.44$ and for two different values of $r=0.47$ (red symbols) and $r=1.88$ (blue symbols). The continuous and dashed lines are the corresponding theoretical predictions given in Eq.~(\ref{eq_C2_norm_letter}) and the symbols represent the experimental data. The error bars are only due to the dispersion between the different beads.}
	\label{fig:std_C2_new}
\end{figure}

We then computed the order statistics $P(M_K,N)$, i.e. the distribution of the $K$-th maximum, for different sets of parameters on a rather large intervals.  
\begin{figure}
 	\centering
        \includegraphics[width=0.99\linewidth]{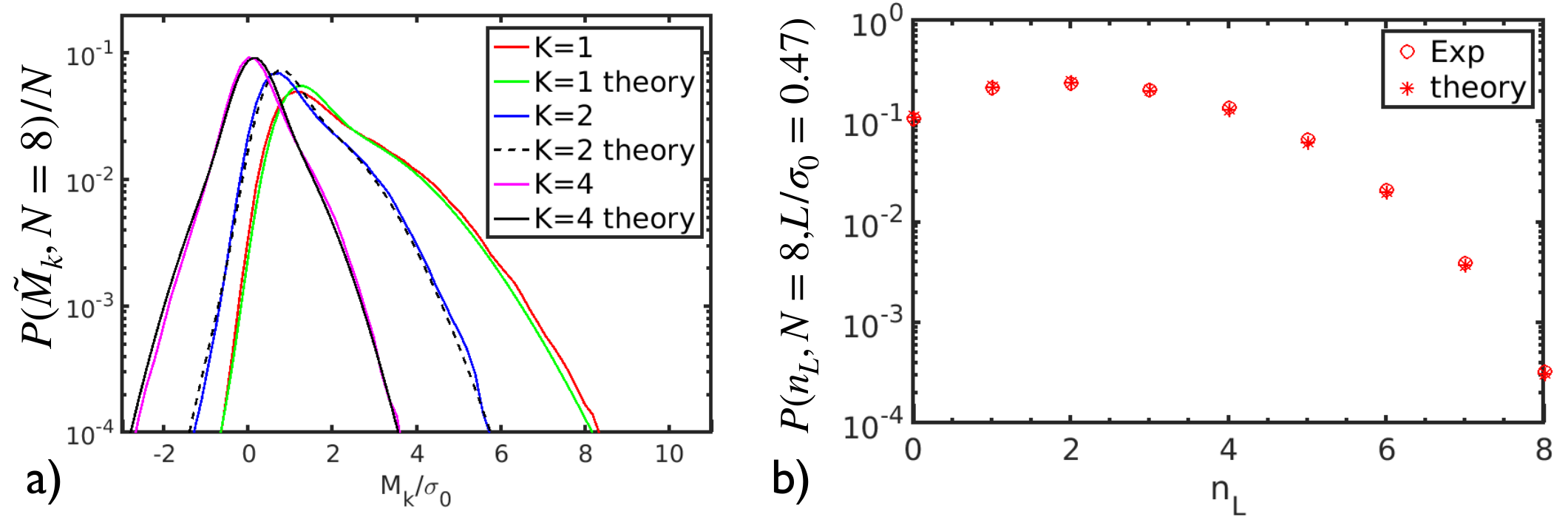}
 	\caption{{\bf a)}: the order statistics $P(\tilde M_K,\ser{N=N_{\rm eff}=8})$ plotted as a function of the scaled value $\tilde M_K = M_K/\sigma_0$. The experimental and theoretical curves are indicated in the legend. The theoretical curves are given by the exact formulae in Eqs. (S7)-(S8) in the \textcolor{black}{SM}. {\bf b)}: the full counting statistics, i.e., the distribution $P(n_L,\ser{N=N_{\rm eff}=8})$  plotted as a function of $n_L$ for fixed dimensionless interval size $L/\sigma_0 \simeq 0.47$. The circles are the experimental values and the stars are the theoretical predictions given in Eqs. (S9)-(S10) in the \textcolor{black}{SM}. Here $k_1 = 1.32$, $k_2=0.17$ and $r=0.188$. See Fig. \ref{fig:pdfMk1} in End Matter for other values of these parameters.}
 	\label{fig:pdfMk1_text}
 \end{figure}
 For the same set of parameters, we also computed the full counting statistics $P(n_L,N)$ denoting the distribution of the number of particles $n_L$ inside a given interval $[-L,L]$ with $L/\sigma_0\simeq 0.47$. The experimental results are compared with the theoretical predictions.  The order statistics $P(\tilde M_K,\ser{N=N_{\rm eff}=8})$ for different values of $K$ and different values of the parameters $k_1, k_2$ and $r$ 
are plotted as a function of $M_K/\sigma_0$ in Fig.~\ref{fig:pdfMk1_text} (a)  and Fig. \ref{fig:pdfMk1},~(a), (c) and (e). The full counting statistics $P(n_L, \ser{N=N_{\rm eff}=8})$ is plotted as a function of $n_L$ in Fig.~\ref{fig:pdfMk1_text}~(b) and Fig.~\ref{fig:pdfMk1}~(b), (d) and (f)   for different $k_1, k_2$, $r$ and for $K=1,2,4$. For all parameter values, the  agreement between the experimental results and the theoretical predictions is excellent taking into account that there are no free parameters. The mean values of $M_1$ are shown in the SM \cite{SM}.

 \paragraph{Conclusions.--}

\textcolor{black}
{To summarize, we have experimentally investigated a system of $N=8$ one dimensional Brownian particles, each confined in their own individual trap whose stiffness switches simultaneously between two values at random times chosen from a Poissonian distribution. Our setup and the protocol eventually drives the system into a non-equilibrium stationary state (NESS) characterized by long-range correlations arising from synchronized back-and-forth switching mechanism. Two main outcomes of our work are the following: (i) if the particles were truly noninteracting, then a stochastic switching will generate nontrivial correlations (DEC). The stochasticity is very important to generate these correlations. 
For example, if the switching was periodic/stroboscopic, we show analytically and numerically (see Section III of \cite{SM}) that no equal-time correlations are generated by such a drive. We also investigated numerically in the $N=2$ particles model how the hydrodynamic interactions between the particles affect the correlations. Surprisingly, we found that for the periodic case, even in the presence of interactions, the correlations are order of magnitude less compared to the stochastic drive. These results are discussed in detail in Section III of the SM \cite{SM}. (ii) The second important finding of this work is that the hydrodynamic interactions clearly are present in the experimental system and, together with the stochastic switching, one arrives at a complex NESS. When the stochastic drive is switched on, some observables, like the correlator $C_1$ defined earlier below Eq. (\ref{eq_C2_norm_letter}), is highly sensitive to the presence of hydrodynamic interactions (see End Matter for details). In contrast, our experiment shows that there are other observables, such as $C_2$ defined in Eq. (\ref{C2}), the statistics of the largest deviation $M_1$ and the full counting statistics that are completely insensitive to the presence of these hydrodynamic interactions (see \cite{SM} for details). This rather surprising experimental observation raises the natural question: why are they so insensitive? 
Furthermore, are there other observables that are also insensitive to these hydrodynamic interactions? These remain important open questions for future work.     
}

\paragraph{Acknowledgments.--}{MB, SNM and GS acknowledge support from ANR Grant No. ANR-23-CE30-0020-01 EDIPS. M. K. acknowledges support from the Department of Atomic Energy, Government of India, under project no. RTI4001.}


\clearpage


\setcounter{equation}{0}
\setcounter{figure}{0}
\renewcommand{\theequation}{EM\arabic{equation}}
\renewcommand{\thefigure}{EM\arabic{figure}}

\section{End matter}

\subsection{Role of Hydrodynamic Interactions} 

\textcolor{black}{
The signature of the hydrodynamic interactions are usually captured by the {\color{black} time-delayed} correlation function 
{\begin{equation}
	C_{i,j}(\delta t) = \frac{\langle x_i(0) x_j(\delta t) \rangle}{\sqrt{\langle x_i^2(0) \rangle \langle x_j^2(0) \rangle}}\, .
	\label{eq:corr}
\end{equation}
Here the argument $0$ refers to the fact that the positions are sampled initially from the stationary state.
}
Hence, when $\delta t=0$, this is the standard correlator $C_1$ defined in the main text below Eq. (\ref{eq_C2_norm_letter}). 
In the absence of the stochastic switching, the {\color{black}time-delayed} correlation function $C_{i,j}(\delta t)$, as a function of $\delta t$, displays a pronounced minimum at a characteristic relaxation time $\tau$ (see the green and the black lines in Fig. \ref{fig:correlation}). Furthermore this equilibrium correlation function $C_{i,j}(\delta t) \to 0$ as $\delta t \to 0$, due to the $x_i \to -x_i$ symmetry of the system \cite{Berut}. In Fig.~\ref{fig:correlation}, for the case of no stochastic switching,  
we show a comparison between the experimental measurement (green) and the theoretical prediction based on a model of two beads~\cite{Berut}. 
The experimental curve (in green) was obtained with $N=4$ particles. 
Despite the fact that there were more than two particles, the model based on two beads provides a rather good description, as evident from Fig. \ref{fig:correlation}. It is natural to ask how this standard signature in $C_{i,j}(\delta t)$ gets altered when the driving via the simultaneous modulation of the traps is switched on. From Fig.~\ref{fig:correlation}, we see that the effect is quite dramatic (purple) and there are two principle features: (i) $C_1 = C_{i,j}(\delta t = 0 )$ becomes a nonzero positive number and (ii) the minimum at $\tau$ disappears. The fact that $C_1>0$ shows the deviation from the noninteracting theory that predicts $C_1=0$ (as discussed in the main text). Had the particles been noninteracting, then, under stochastic switching $C_1$ would still be zero due to the $x_i \to -x_i$ symmetry, as discussed in the main text. The fact that the experimental curve (in purple) shows that $C_1 >0$ clearly indicates that there are nonzero hydrodynamic interactions between the particles. 
The disappearance of the minimum indicates that the stochastic driving significantly reduces the {\color{black}time-delayed} anti-correlation between two beads.} 

\textcolor{black}{Hence we see that the stochastic switch enhances the correlations at $\delta t = 0$, while it reduces the correlations when $\delta t \sim \tau$. 
Therefore, hydrodynamic interactions
and the stochastic switching interplay to create a NESS with complex nontrivial correlations between the particles. Clearly the observable $C_1 = C_{i,j}(\delta t =0)$ is sensitive to the presence of hydrodynamic interactions. However, the other two observables discussed in the main text, namely the correlator $C_2$ and the distribution of the position $M_1$ of the rightmost particle, seem to be insensitive to the presence of hydrodynamic interactions. This is
our surprising experimental observation. A natural question would be to investigate whether there are other observables that are also insensitive to the presence of hydrodynamic interactions.
\vspace{0.2cm}
\begin{figure}
	\centering
\includegraphics[width=\linewidth]{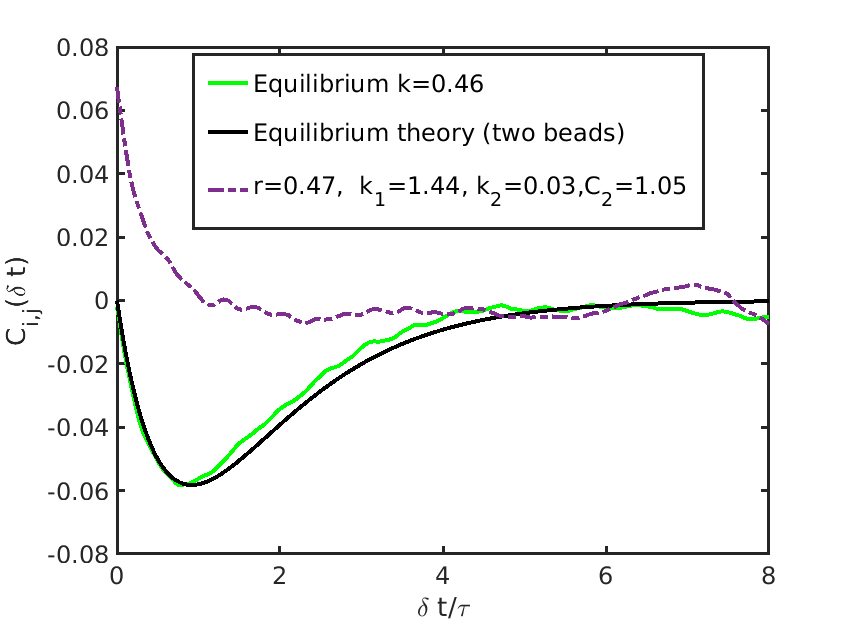}
\caption{{\color{black} Time-delayed correlation $C_{i,j}(\delta t)$ defined in Eq.~\eqref{eq:corr}, plotted as a function of $\delta t$. We see that, in the absence of a stochastic switching, i.e., in equilibrium, the experimental data (green) is in good agreement with the theoretical prediction (black) based on two hydrodynamically coupled particles~\cite{Berut}. When the stochastic drive is switched on, the purple curve describes the experimental data and shows a ``creation'' of nonzero positive correlation at $\delta t = 0$ and a suppression of anticorrelation near the minimum $\delta t \sim \tau$.} }
	\label{fig:correlation}
\end{figure}
}

\subsection{Statistics of the largest deviation $M_1$ and the full counting statistics}

\textcolor{black}{
The distribution of the adimensional maximum, i.e., the {scaled} position of the rightmost particle {$\tilde{M}_1 =\max \{z_i\}$} is given by~\cite{BKMS24}
{
\begin{align}
    P(\tilde{M}_1,N) = N &\int_0^{1} \dd u\; h(u) \frac{e^{-\frac{\tilde{M}_1^2}{2 V(u)}}}{\sqrt{2 \pi V(u)}} \nonumber \\
    &\times \left[ \frac{1}{2} \left( 1 + {\rm erf}\left(\frac{\tilde{M}_1}{\sqrt{2 V(u)}}\right) \right) \right]^{N-1} \;, \label{eq:max}
\end{align}
}
where we recall that $V(u) = \left[\frac{u}{k_2} + \frac{1 - u}{k_1}\right]$ and 
\begin{equation} \label{eq:def-h_EM}
    h(u) = A \, u^{R_1 - 1} (1-u)^{R_2 - 1} V(u) \;,
\end{equation}
with $R_i = \frac{r_i}{2 k_i}$ with $i=1,2$. For a derivation of this result, see Eq. (S7) in the SM \cite{SM}. 
}

\textcolor{black}{In addition, we present further experimental data (complementing the ones shown in Fig. 3 in the main Letter) and compare them to the theoretical results for two observables: (i) the distribution of the dimensionless position of the $K$-th particle $\tilde M_{K}$, i.e,. the $K$-th particle counting from the right (order statistics), (ii) the full counting statistics (FCS) $n_L$, denoting the number of components (both $x_i$ and $y_i$) that are inside the interval $[-L,+L]$ around the trap center.}    

\begin{figure}
 	\centering
 	{\bf \hspace{-3cm}a) \hspace{4cm} b)} \\
 		\includegraphics[width=0.48\linewidth]{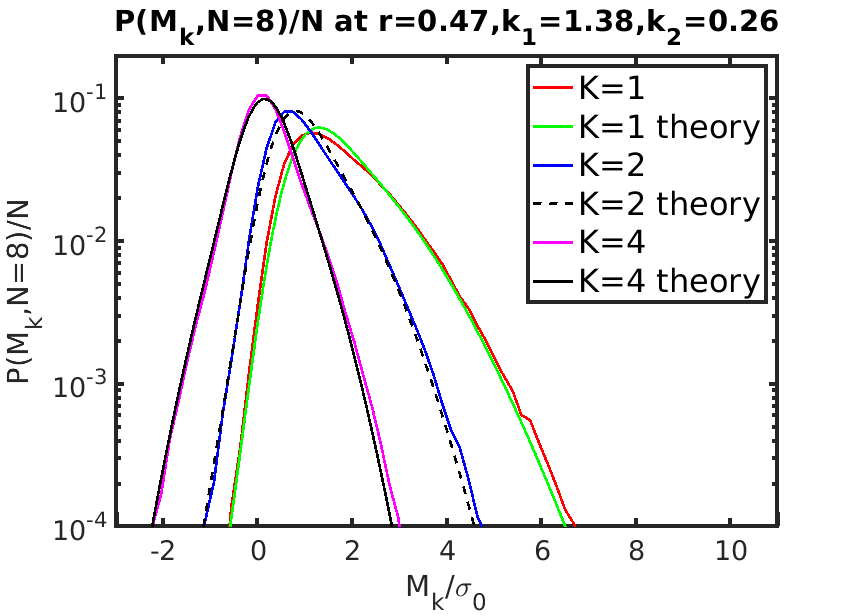}
 		\includegraphics[width=0.48\linewidth]{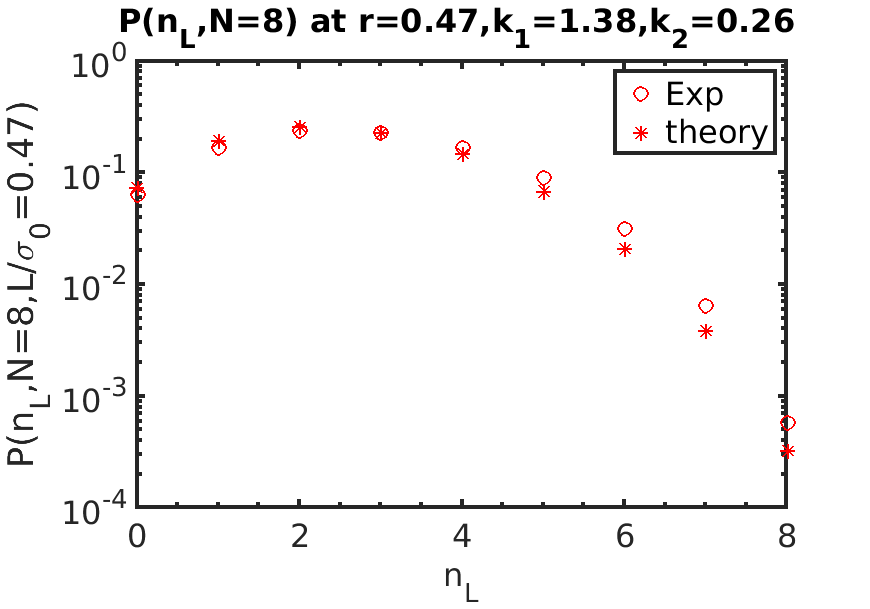} \\
 		{\bf \hspace{-3cm}c) \hspace{4cm} d)} \\
 		\includegraphics[width=0.48\linewidth]{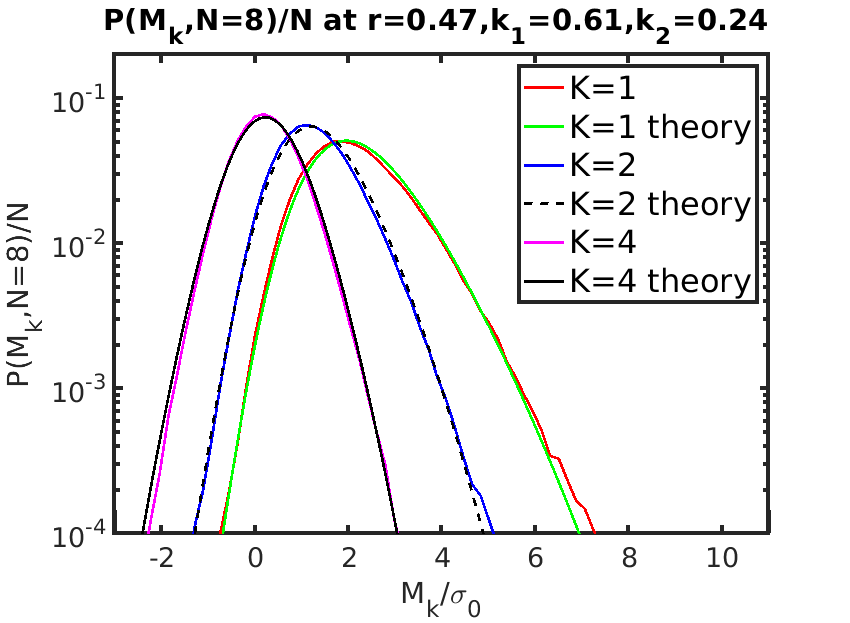}
		   \includegraphics[width=0.48\linewidth]{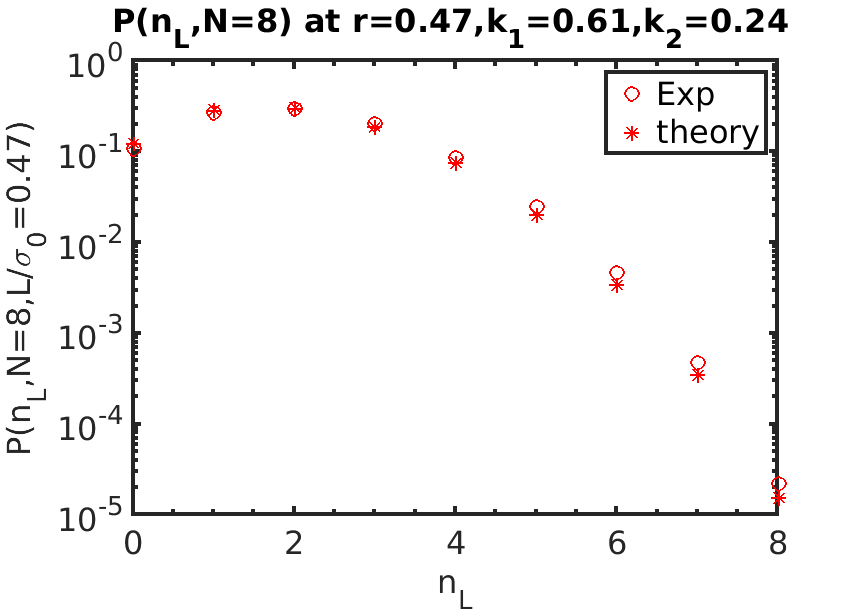}	\\
		{\bf \hspace{-3cm}e) \hspace{4cm} f)} \\
		\includegraphics[width=0.48\linewidth]{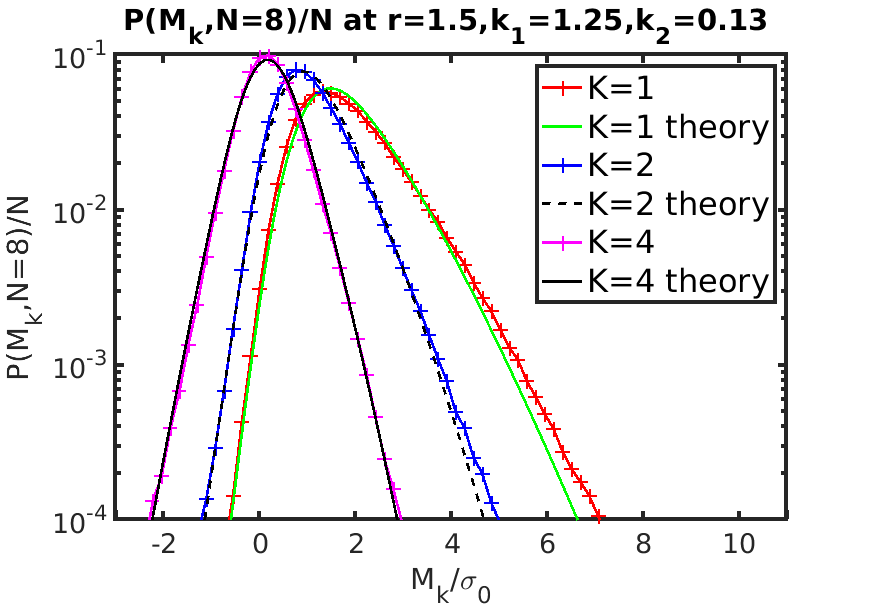}
		\includegraphics[width=0.48\linewidth]{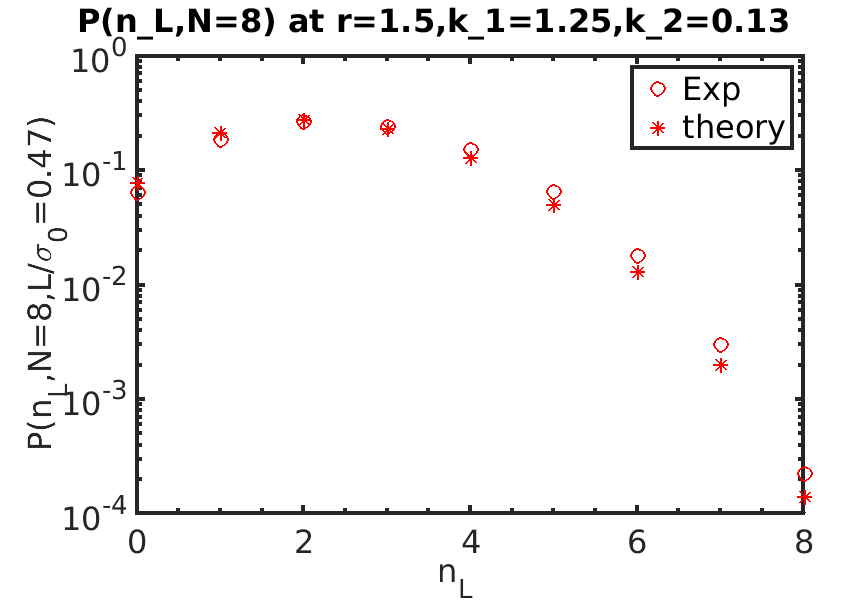}
 	\caption{Panel (a),(c),(e): the order statistics $P(\tilde M_K,\ser{N=N_{\rm eff}=8})$ plotted as a function of the scaled value $\tilde M_K = M_K/\sigma_0$ for different values of $k_1, k_2$ and $r$. The experimental and theoretical curves are indicated in the legend. The theoretical curves are given by the exact formulae in Eqs. (S7)-(S8) in the \textcolor{black}{SM}. Panel (b),(d),(f): the full counting statistics, i.e., the distribution $P(n_L,\ser{N=N_{\rm eff}=8})$  plotted as a function of $n_L$ for fixed dimensionless interval size $L/\sigma_0 \simeq 0.47$. The circles are the experimental values and the stars are the theoretical predictions given in Eqs. (S9)-(S10) in the \textcolor{black}{SM}.}
 	\label{fig:pdfMk1}
 \end{figure}
 
\newpage


\begin{thebibliography}{99}

\bibitem{huygens_book}
J. G. Yoder,  {\it Christiaan Huygens, book on the pendulum clock (1673)}. In {\it Landmark Writings in Western Mathematics}, 1640-1940 (pp. 33-45). Elsevier Science (2005).


\bibitem{NBM05} A. Nagar, M. Barma, S.~N. Majumdar, {\it Passive Sliders on Fluctuating Surfaces:
Strong-Clustering States}, Phys. Rev. Lett. {\bf 94}, 240601 (2005).

\bibitem{NBM06} A. Nagar, S.~N. Majumdar, M. Barma, {\it Strong clustering of non-interacting, passive
sliders driven by a Kardar-Parisi-Zhang surface}, Phys. Rev. E {\bf 74}, 021124 (2006).


\bibitem{MBMS20}
G. Mercado-Vasquez, D. Boyer, S. N. Majumdar, G. Schehr, {\it Intermittent resetting potentials}, J. Stat. Mech., 113203 (2020).

\bibitem{GK21}
D. Gupta, A. Kundu, {\it Resetting with stochastic return through linear confining potential}, J. Stat. Mech., 043202 (2021).

\bibitem{SDN21}
I. Santra, S. Das, S. K. Nath, {\it Brownian motion under intermittent harmonic potentials}, J. Phys. A: Math. Theor. {\bf 54}, 334001 (2021).


\bibitem{MBM22}
G. Mercado-Vasquez, D. Boyer, S. N. Majumdar, {\it Reducing mean first passage times with intermittent confining potentials: a realization of resetting processes}, J. Stat. Mech., 093202 (2022).

\bibitem{BKMS24} M. Biroli, M. Kulkarni, S. N. Majumdar, G. Schehr, {\it Dynamically emergent correlations between particles in a switching harmonic trap}, Phys. Rev. E {\bf 109}, L032106 (2024).


\bibitem{BLMS23} M. Biroli, H. Larralde, S. N. Majumdar, and G. Schehr, {\it Extreme statistics and spacing distribution in a Brownian gas correlated by resetting}, Phys. Rev. Lett. {\bf 130}, 207101 (2023).

\bibitem{BLMS24} M. Biroli, H. Larralde, S. N. Majumdar, G. Schehr, {\it Exact extreme, order, and sum statistics in a class of strongly correlated systems}, Phys. Rev. E {\bf 109}, 014101 (2024).


\bibitem{SM2024}
S. Sabhapandit, S. N. Majumdar, {\it Noninteracting particles in a harmonic trap with a stochastically driven center}, J. Phys. A: Math. Theor. {\bf 57}, 335003 (2024).

\bibitem{KMS2025}
M. Kulkarni, S. N. Majumdar, S. Sabhapandit, {\it Dynamically emergent correlations in bosons via quantum resetting}, J. Phys. A: Math. Theor. {\bf 58}, 105003 (2025).


\bibitem{MMS2025}
N. Mesquita, S. N. Majumdar, S. Sabhapandit, {\it Dynamically emergent correlations in a Brownian gas with diffusing diffusivity}, preprint arXiv:2506.20859 (2025).





\bibitem{Berut} A. Berut, A. Petrosyan, S. Ciliberto {\it Energy flow between two
hydrodynamically coupled particles kept at different effective temperatures}, Europhys. Lett. {\bf 107}, 60004 (2014).


\bibitem{kotar} J. Kotar, M. Leoni, B. Bassetti, M. C. Lagomarsino, P. Cicuta, {\it Hydrodynamic synchronization of colloidal oscillators}, Proc. Natl. Acad. Sci. USA, \textbf{107}, 7669 (2010).

\bibitem{Barlett} P. Bartlett, S. I. Henderson, S. J. Mitchell, {\it Measurement of the hydrodynamic forces between two polymer coated spheres}, Philos. Trans. R. Soc. Lond. Ser. A, \textbf{359}, 883 (2011).


\bibitem{polin}
{M. Polin, D. G. Grier, S. R. Quake, {\it Anomalous vibrational dispersion in holographically trapped colloidal
arrays}, Phys. Rev. Lett. {\bf 96}, 088101 (2006).}

\bibitem{gao} L. Gao, N. J. Gottron III, L. N. Virgin, B. B. Yellen, {\it The synchronization of superparamagnetic beads driven by a micro-magnetic ratchet}, Lab Chip, \textbf{10}, 2108 (2010).

\bibitem{dobnikar} J. Dobnikar, A. Snezhko, A. Yethiraj, {\it Emergent colloidal dynamics in electromagnetic fields}, Soft Matter, \textbf{9}, 3693 (2013).

\bibitem{yan} J. Yan, M. Bloom, S. C. Bae, E.  Luijten, S. Granick, {\it Linking synchronization to self-assembly using magnetic, Janus colloids}, Nature, \textbf{491}, 578 (2012).

\bibitem{ignacio} I. A. Martinez, C. Devailly, A. Petrosyan, S. Ciliberto, {\it Energy Transfer between Colloids via Critical Interactions}, Entropy, 19, No 2, (2017).





\bibitem{EM11a}
M. R. Evans, S. N. Majumdar, {\it Diffusion with stochastic resetting}, Phys. Rev. Lett. {\bf 106}, 160601 (2011).

\bibitem{EM11b}
M. R. Evans, S. N. Majumdar, {\it Diffusion with Optimal Resetting}, J. Phys. A: Math. Theor. {\bf 44}, 435001 (2011).




\bibitem{EMS20}
M. R. Evans, S. N. Majumdar, G. Schehr, {\it Stochastic resetting and applications}, J. Phys. A: Math. Theor. {\bf 53}, 193001 (2020).


\bibitem{PKR22}
A. Pal, S. Kostinski, S. Reuveni, {\it The inspection paradox in stochastic resetting}, J. Phys. A: Math. Theor. {\bf 55}, 021001 (2022).

\bibitem{GJ22}
S. Gupta, and A. M. Jayannavar, {\it Stochastic resetting: A (very) brief review}, Front. Phys. {\bf 10} 789097, (2022).




\bibitem{Roichman20}
O. Tal-Friedman, A. Pal, A. Sekhon, S. Reuveni, Y. Roichman, {\it Experimental realization of diffusion with stochastic resetting}, J. Phys. Chem. Lett. {\bf 11}, 7350 (2020).

\bibitem{BBPMC_20} B. Besga, A. Bovon, A. Petrosyan, S. N. Majumdar, S. Ciliberto, {\it Optimal mean first-passage time for a Brownian searcher subjected to resetting: experimental and theoretical results}, Phys. Rev. Research {\bf 2}, 032029(R) (2020).




\bibitem{FBPCM_21} F. Faisant, B. Besga, A. Petrosyan, S. Ciliberto, S. N. Majumdar, {\it Optimal mean first-passage time of a Brownian searcher with resetting in one and two dimensions: experiments, theory and numerical tests}, J. Stat. Mech., 113203 (2021).


\bibitem{Landauer23}
R. Goerlich, M. Li, L. B. Pires, P. A. Hervieux, G. Manfredi, C. Genet, {\it Taming a Maxwell's demon for experimental stochastic resetting}, preprint arXiv:2306.09503 (2023).



\bibitem{DHP14}
X. Durang, M. Henkel, H. Park, J. Phys. A: Math. Theor. {\bf 47}, 045002 (2014).


\bibitem{GMS14}
S. Gupta, S. N. Majumdar, G. Schehr, {\it Fluctuating interfaces subject to stochastic resetting}, Phys. Rev. Lett. {\bf 112}, 220601 (2014).

\bibitem{BBR16}
U. Bhat, C. De Bacco, S. Redner, {\it Stochastic search with Poisson and deterministic resetting}, J. Stat. Mech. 083401, (2016).

\bibitem{MSS18}
B. Mukherjee, K. Sengupta, S. N. Majumdar, {\it Quantum dynamics with stochastic reset}, Phys. Rev. B {\bf 98}, 104309 (2018). 

\bibitem{RTLG18}
D. C. Rose, H. Touchette, I. Lesanovsky, J. P. Garrahan, {\it Spectral properties of simple classical and quantum reset processes}, Phys.
Rev. E {\bf 98}, 022129 (2018).



\bibitem{BKP19}
U. Basu, A. Kundu, A. Pal., {\it Symmetric exclusion process under stochastic resetting}, Phys. Rev. E {\bf 100}, 032136 (2019).
\bibitem{MMS20}
M. Magoni, S. N. Majumdar, G. Schehr, {\it Ising model with stochastic resetting}, Phys. Rev. Res. {\bf 2}, 033182 (2020).





\bibitem{PCPL21}
G. Perfetto, F. Carollo, M. Magoni, L. Lesanovsky, {\it Designing nonequilibrium states of quantum matter through stochastic resetting}, Phys. Rev. B {\bf 104}, L180302 (2021).

\bibitem{NG23} A. Nagar and S. Gupta, {\it Stochastic resetting in interacting particle systems: A review}, J. Phys. A: Mathematical and Theoretical {\bf 56}, 28 (2023).


\bibitem{BMS23} M. Biroli, S. N. Majumdar, G. Schehr, {\it Critical number of walkers for diffusive search processes with resetting}, Phys. Rev. E {\bf 107}, 064141 (2023).


\bibitem{BMS25}
M. Biroli, S. N. Majumdar, G. Schehr, {\it Resetting Dyson Brownian motion}, Phys. Rev. E {\bf 112}, 014101 (2025).














\bibitem{VR25} R. Vatash, Y. Roichman, {\it Many-Body Colloidal Dynamics under Stochastic Resetting: Competing Effects of Particle Interactions on the Steady State Distribution}, preprint arXiv:2504.10015.

\bibitem{SM} M. Biroli, S. Ciliberto, M. Kulkarni, S. N. Majumdar, A. Petrosyan, G. Schehr, see Supplementary Material which contains references [45-56] for more details.

\bibitem{MS24} S.~N. Majumdar, G. Schehr, {\it Statistics of Extremes and Records in Random Sequences}, Oxford University Press, (2024).


   
\bibitem{ref:Quake} 
J.-C. Meiners, S. R. Quake, 
{\it Direct Measurement of Hydrodynamic Cross Correlations between Two Particles in an External Potential}, {Phys. Rev. Lett.}, {\bf 82}, 2211 ({1999}).

	
\bibitem{ref:Ou-Yang}
L. A. Hough, H. D. Ou-Yang, {\it Correlated motions of two hydrodynamically coupled particles confined in separate quadratic potential wells},
{Phys. Rev. E} {\bf 65}, {021906}({2002}). 
	


\bibitem{ref:Herrera}
	S. Herrera-Velarde, E. C. Eu\'{a}n-D\'{i}az, F. C\'{o}rdoba-Vald\'{e}s, R. Casta\~{n}eda-Priego, {\it Hydrodynamic correlations in three-particle colloidal systems in harmonic traps}, {J. Phys.: Condens. Matter} {\bf 25}, {325102} ({2013}).
    
	
\bibitem{Baiesi} C. Yolcu, A. Berut, G. Falasco, A. Petrosyan, S. Ciliberto, and M. Baiesi, {\it A General Fluctuation-Response Relation for Noise Variations and its Application to Driven Hydrodynamic
Experiments}, J. Stat. Phys. {\bf 167}, 29 (2017).
	
	\bibitem{Berut_2016} A. Berut, A. Imparato, A. Petrosyan, S. Ciliberto, {\it The role
	of coupling on the statistical properties of the energy fluxes between
	stochastic systems at different temperatures}, J. Stat. Mech. 054002 (2016).
	
	\bibitem{Berut_2016B} A. Berut, A. Imparato, A. Petrosyan, S. Ciliberto,  {\it Stationary and
	Transient Fluctuation Theorems for Effective Heat Fluxes between Hydrodynamically Coupled Particles in Optical Traps}, Phys. Rev. Lett. {\bf 116}, 068301 (2016).
	
	\bibitem{Martinez_2023} I. A. Martinez, A. Petrosyan, S. Ciliberto, {\it Laser-
	induced heating for the experimental study of critical Casimir forces
	with optical trapping}, SciPost Phys. {\bf 15}, 247 (2023).
	
	\bibitem{Grier} J. Crocker, D. Grier, {\it Methods of Digital Video Microscopy for Colloidal Studies}, J. Colloid Interface Sci. {\bf 179}, 298 (1996). 



   
	
\bibitem{Ashkin_2006} A. Ashkin, {\it Optical Trapping and Manipulation of Neutral Particles Using Lasers: A Reprint Volume with Commentaries}, (World Scientific, 2006).

\bibitem{Volpe} P. H. Jones, O. M. Marag\'o, G. Volpe, {\it Optical Tweezers principle and applications}, Cambridge University Press (2015).
	
\bibitem{Gittes} F. Gittes, C. F. Schmidt,
	{\it Signals and noise in micromechanical measurements}, Methods Cell Biol. {\bf 55}, 129 (1998). 
	
	
\bibitem{Capitanio}	M. Capitanio, G. Romano, R. Ballerini, M. Giuntini, F. S. Pavone, {\it Calibration of optical tweezers with differential interference
	contrast signals}, Rev. Sci. Instrum. {\bf 73}, 1687 (2002).

































\end{thebibliography}
\end{document}


\title{Experimental evidence for strong emergent correlations between particles in a switching trap}

\author{Marco Biroli}
\affiliation{LPTMS, CNRS, Univ.  Paris-Sud,  Universit\'e Paris-Saclay,  91405 Orsay,  France}
\author{Sergio Ciliberto}
\affiliation{Laboratoire de Physique, CNRS, ENS de Lyon, F-69342 Lyon, France}
\author{Manas Kulkarni}
\affiliation{ICTS, Tata Institute of Fundamental Research, 560089 Bengaluru, India}
\author{Satya N. Majumdar}
\affiliation{LPTMS, CNRS, Univ.  Paris-Sud,  Universit\'e Paris-Saclay,  91405 Orsay,  France}
\author{Artyom Petrosyan}
\affiliation{Laboratoire de Physique, CNRS, ENS de Lyon, F-69342 Lyon, France}
\author{Gr\'egory Schehr}
\affiliation{Laboratoire de Physique Th\'eorique et Hautes Energies, CNRS UMR 7589, Sorbonne Universit\'e, 4 Place Jussieu, 75252 Paris Cedex 05, France}

\date{\today}

\begin{abstract}
    In this Supplemental Material, we provide details relevant for the results given in the main text.
\end{abstract}

\maketitle

\thispagestyle{fancy}


\tableofcontents

\section{Details of the analytical results presented in the Letter}

\subsection{Normalization constant}

The normalization constant of Eq.~(2) in the main text which was omitted there for brevity is given by
\begin{equation}
    A = \frac{r_1 k_2 + r_2 k_1}{r_1 + r_2} B\left( \frac{r_1}{2 k_1}, \frac{r_2}{2 k_2} \right) \;,
\end{equation}
where $B(m,n)$ is the standard Beta function. {We recall here that $k_1$,  $k_2$, $r_1$, $r_2$ are all measured in adimensional units. }

\subsection{Derivation of main observables}

We provide a quick overview of the derivations involved in computing the analytical formulas for the observables studied in this Letter. Our starting point is Eq.~(1) and Eq.~(2) in the main text. As stated in the main Letter, exploiting the CIID structure of the JPDF in Eq.~(1) in the main text, we first fix the conditioning variable $u$. Once $u$ is fixed, the variables are IID and calculating these observables for such IID variables is easy. Finally, these observables for fixed $u$ are averaged over all possible values of $u$ drawn from the PDF $h(u)$. These results were derived in Ref.~\cite{BKMS24}. Here we just briefly recapitulate them, for the purpose of completeness and to adapt to the notations of this paper. We will now systematically look at the connected {two-point} correlator, the maximum, the order statistics and the full counting statistics. 

\vspace{0.2cm}

\noindent{\bf The variance.} From the JPDF in Eq.~(1) of the main text, the variance $\langle z_i^2\rangle$ can be easily computed \cite{BKMS24} and in adimensional units it reads
\begin{equation}
{\rm Var}(z_i) = \langle z_i^2\rangle = \int_0^1 du\, h(u)\, V(u) \;,
\end{equation}
where $h(u)$ and $V(u)$ are defined in Eq.~(2) in the main text. Performing this integral yields the result in Eq.~(3) of the main text.

\vspace{0.2cm}
\noindent{\bf The connected {two-point} correlator.} The first non-zero {two-point} correlator of the JPDF in Eq.~(1) in the main text is
\begin{equation}
    \tilde C_{2} = \langle z_i^2 z_j^2 \rangle - \langle z_i^2 \rangle \langle z_j^2 \rangle \;,
\end{equation}
for $i \neq j$. We compute it from the JPDF in Eq.~(1) of the main text and average over $u$ to get
\begin{equation}
   \tilde C_2 = \int_0^{1} \dd u \; h(u) V(u)^2 - \left[\int_0^{1} \dd u \; h(u) V(u) \right]^2 \;.
\end{equation}
Replacing $h(u)$ with its definition in Eq.~(2) in the main text as well as $V(u) = \left[\frac{u}{k_2} + \frac{1 - u}{k_1}\right]$ we can compute these integrals explicitly. The full formula for any $r_1, r_2$ is quite lengthy and {given in the Supplementary Material of Ref.~\cite{BKMS24}}. Experimentally we focus on the case where $r_1 = r_2 = r$, for which the expression simplifies considerably 
{\begin{equation}
  \tilde  C_2 = \frac{(R_1 - R_2)^2 (2 + 3 R_1 + 3 R_2 + 4 R_1 R_2) }{(1 + R_1 + R_2)^2 (2 + R_1 + R_2)} \;,
\end{equation}}
where we introduced the dimensionless constants $R_i = r / (2 k_i)$ for brevity. Similarly the normalized correlator
\begin{equation}
C_{2}=\frac{\langle z_i^2 z_j^2 \rangle}{\langle z_i^2 \rangle \langle z_j^2 \rangle}-1
\end{equation}
is given by Eq.~(5) in the main text. 
%

\vspace{0.2cm}

\noindent{\bf The maximum and the order statistics.} We start by looking at the dimensionless position of the $K$-th particle $\tilde M_{K}$, i.e,. the $K$-th particle counting from the right. The maximum corresponds to the special case $K = 1$. Since the JPDF of the scaled positions of the particles $P(z_i)$
 in Eq.~(1) in the main text has a CIID structure with $u$ as a conditioning variable (drawn from the distribution $h(u)$), we first compute the order statistics of the random variables with a fixed $u$ and then average over $h(u)$. For fixed $u$, one sees from Eq.~(1) in the main text that the $z_i$ variables are zero mean independent Gaussian variables with variance $V(u)$. We denote the scaled $K$-th maximum for fixed $u$ as $\tilde M_K(u)$. The distribution of $\tilde M_K(u)$ can thus be expressed as  
{\begin{align}
    P(\tilde{M}_K(u),N) &= (N - K + 1) \binom{N}{K - 1} \frac{e^{-\frac{\tilde{M}_L(u)^2}{2 V(u)}}}{\sqrt{2 \pi V(u)}} \left[ \frac{1}{2} \left( 1 + {\rm erf}\left( \frac{\tilde{M}_K(u)}{\sqrt{2 V(u)}} \right) \right) \right]^{N - K} \left[ \frac{1}{2} \, {\rm erfc}\left(\frac{\tilde{M}_K(u)}{\sqrt{2 V(u)}}\right)\right]^{K - 1} \label{eq:order3} \;,
\end{align}}
where ${\rm erf}(z) = (2/\sqrt{\pi}) \int_0^z e^{-u^2}\,du$ is the error function and ${\rm erfc}(z) = 1 - {\rm erf}(z)$ is the complementary error function. The different terms on the right hand side (RHS) of this result can be understood as follows. For the $K$-th scaled maximum to be located at $\tilde M_K(u)$ we must choose $K - 1$ particles to be above $\tilde M_K(u)$, accounting for the binomial term in Eq.~(\ref{eq:order3}). Subsequently, we need to choose one of the remaining $N - K + 1$ particles to place at $\tilde M_{K}(u)$, accounting for the remaining two terms in Eq.~(\ref{eq:order3}). The last $N - k$ particles must be placed below $\tilde M_{K}(u)$, which corresponds to Eq.~(\ref{eq:order3}). Using Eq.~(1) in the main text, we can write the order statistics $\tilde M_K$ in the steady state of the switching trap as a function of $\tilde M_K(u)$ as
\begin{equation} \label{eq:order-full}
    P(\tilde M_K,N) = \int_0^{1} \dd u\; h(u) P(\tilde M_K(u),N) \;.
\end{equation}
{Although} this integral cannot be computed explicitly we can evaluate it numerically for any value of $N, k_1, k_2, r_1, r_2$, , allowing us to compare our experimental results to this exact theoretical expression.

\noindent{\bf Full counting statistics. } The full counting statistics is defined as the distribution $P(n_L,N)$ of the number of particles 
$n_L$ in an interval $[-L, L]$ around the trap center. 
Using again the CIID structure of the JPDF in Eq.~(1) in the main text, we first fix the conditioning variable $u$ and denote by $n_L(u)$ the number of particles in $[-L,L]$ for fixed $u$. Using the fact that, for fixed $u$, the variables $z_i$'s are zero mean independent Gaussians with variance $V(u)$, the full counting statistics for fixed $u$ can be written as
\begin{equation}
    P(n_L(u),N) = \binom{N}{n_L(u)} {\rm erf}\left(\frac{L}{\sqrt{2 V(u)}}\right)^{n_L(u)}  \left[ 1 - {\rm erf}\left(\frac{L}{\sqrt{2 V(u)}}\right)\right]^{N - n_L(u)}  \label{eq:fcs2}\;,
\end{equation}
which can be understood as follows. The probability for a centered Gaussian variable of variance $V(u)$ to be in $[-L, L]$ is given by ${\rm erf}(L/\sqrt{2 V(u)})$. Then the probability of $n_L(u)$ particles being in $[-L, L]$ is a binomial distribution where a success corresponds to a particle being in $[-L, L]$, which yields Eq.~(\ref{eq:fcs2}). Once again, exploiting the CIID structure of Eq.~(1) in the main text we can relate the full counting statistics $N_L$ in the steady state of the switching harmonic trap to $n_L(u)$ via
\begin{equation}
    P(n_L,N) = \int_0^{1} \dd u\; h(u) P(n_L(u),N) \;. \label{eq:fcs-full}
\end{equation}
Although Eq.~(\ref{eq:fcs-full}) cannot be evaluated explicitly, we can numerically integrate it for any value of $N, k_1, k_2, r_1, r_2$, allowing us to compare our experimental results to this exact theoretical expression.
\vspace{0.2cm}

\vspace{0.2cm}

\vspace{0.2cm}
\begin{figure}[h!]
	\centering
	{\bf \hspace{-3cm}a) \hspace{4cm} b)} \\ 
	\includegraphics[width=0.45\linewidth, angle=0]{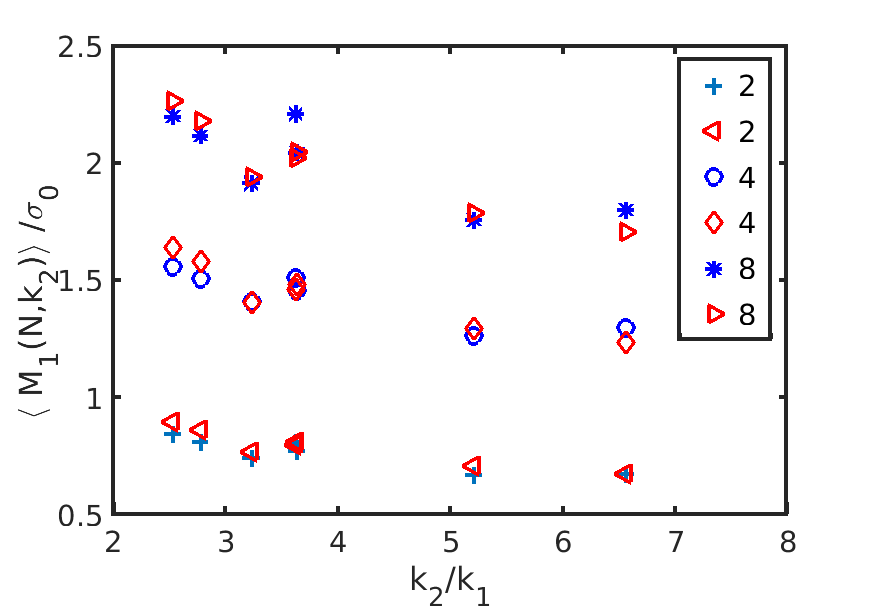}
	\includegraphics[width=0.45\linewidth, angle=0]{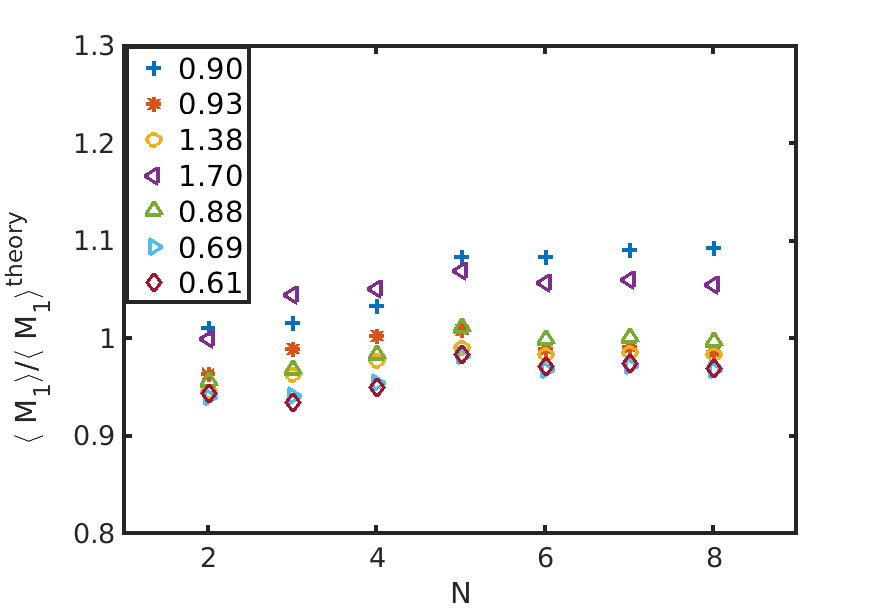} 
	\caption{ (a) The measured  $\langle M_1(N)\rangle$ (blue symbols) and the theoretical predictions $\langle M_1\rangle^{\rm theory}$ (red symbols) are plotted as a function of $k_2/k_1$ for different $N$ (symbols in the legend) at $k_1=0.25$ and $r=0.47$. (b) The ratio between the measured and the theoretical values of $\langle M_1(N)\rangle$ are plotted as a function of $N$ for different values of $k_2$ (symbols in the legend) at $k_1=0.25$ and $r=0.47$. }
	\label{fig:M1_v_k2}
\end{figure}

\noindent{\bf The values of $\langle M_1(N )\rangle$.} In order to strengthen the comparison of the experimental results with the theoretical ones we also computed $\langle M_1(N )\rangle$ versus $N, k_1$ and $k_2$ for different values of $r$. In Fig.~\ref{fig:M1_v_k2} we plot the results at $r = 0.47$. The
errors are always smaller than $10\%$ which is rather good
taking into account that there are no free parameters.

\section{Numerical simulations on the effects on correlations of the hydrodynamic coupling} \label{Sec_hyd_coup}
To understand the effect of the hydrodynamic coupling and to check for possible experimental artifacts, we can use the classical hydrodynamic interaction between two particles. Following \cite{ref:Quake,ref:Bartlett,ref:Ou-Yang} the motion of two identical particles of radius $R$ trapped at positions $X_{1,0}$ and $X_{2,0}$ separated by a distance $d = X_{2,0} - X_{1,0}$, is described by two coupled Langevin equations
\begin{equation}
\label{eq:hydro}
	\left( \begin{array}{c} \dot{x}_1 \\ \dot{x}_2 \end{array} \right) = 
	\mathcal{H}
	\left( \begin{array}{c} F_1 \\ F_2 \end{array} \right)
\end{equation}
where $x_i(t)$ is the displacement of the $i$-th particle relative to its trapping position. In Eq.~\eqref{eq:hydro},  $\mathcal{H}$ is a $2\times 2$ matrix encoding the hydrodynamic coupling and $F_i$ is the force acting on the particle $i$. In the case where the displacements are small compared to the mean distance between the particles, the matrix $\mathcal{H}$ can be expressed as 
\begin{equation}
\label{eq:hmat}
	\mathcal{H} = 
	\begin{pmatrix} 1/\gamma & \epsilon/\gamma \\
		\epsilon/\gamma & 1/\gamma \end{pmatrix} \;,
\end{equation}
where $\gamma$ is the Stokes friction coefficient ($\gamma = 6 \pi R \eta$ with $\eta$ denoting the viscosity of water) and $\epsilon$ is the hydrodynamic coupling coefficient given by $\epsilon = \frac{3R}{2d} - \left(\frac{R}{d}\right)^3$ (as given by the Rotne-Prager diffusion form \cite{ref:Herrera}).

At equilibrium the forces acting on the particles are
\begin{equation}
\label{eq:force}
	F_i = -k(t)\,  x_i + f_i \;,
\end{equation}
where $k(t)$ represents the stiffness of the trap at time $t$. We consider the case when $k(t)$ switches between two values $k_1$ and $k_2$ dichotomously with switching rate $r$. This implies that the intervals between two successive switchings are distributed independently and each is drawn from the probability density function $\psi(\tau)= r\,e^{-r\,t}$. The noise $f_i$'s in Eq.~\eqref{eq:force} are Gaussian random forces that satisfy
\begin{equation}
	\begin{array}{c}
		\langle f_i(t) \rangle = 0 \\
		\langle f_i(t) f_j(t^{\prime}) \rangle = 2k_{\mathrm{B}}T \, (\mathcal{H}^{-1})_{ij} \, \delta(t-t^{\prime})\,.
	\end{array}
\end{equation}
Here $k_{\mathrm{B}}$ is the Boltzmann constant and $T$ the temperature of the surrounding fluid.

Substituting Eqs.~\eqref{eq:hmat} and~\eqref{eq:force} in Eq.~\eqref{eq:hydro} leads to the coupled Langevin equations 
\begin{equation}
	\left\{
	\begin{array}{l}
		\gamma \dot{x}_1 = -k(t) x_1 + \epsilon (-k(t) x_2 + f_2) + f_1 \\
		\gamma \dot{x}_2 = -k(t) x_2 + \epsilon (-k(t) x_1 + f_1 ) + f_2\,. 
	\end{array}
	\right.
	\label{eq:coupledLangevin}
\end{equation}

\begin{figure}
	\includegraphics[width=0.8\columnwidth]{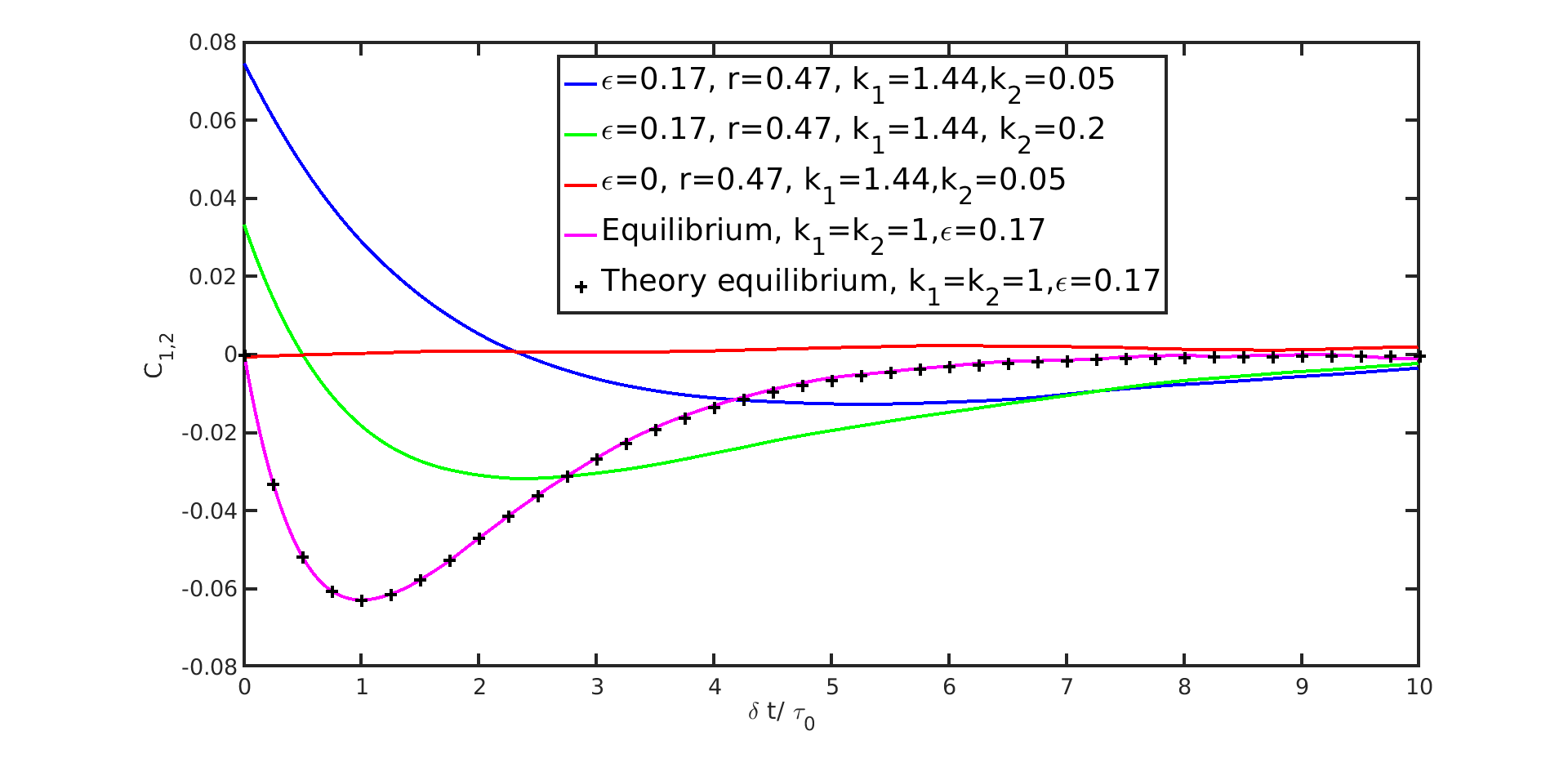}
	\caption{The {cross-correlation} function $C_{1,2}(\delta t)$, defined in Eq.~\eqref{eq:corr}, is plotted versus the delay time $\delta t$ for different values of the forcing with and without coupling. The equilibrium case with $k_1=k_2=1$ is also plotted, {(the magenta line is the numerical result while the black symbols indicate the analytical results from~\cite{Berut}}. When $\epsilon=0$ there is no coupling and the correlation is zero for all forcing. In contrast in the presence of coupling $\epsilon=0.17$ we see that the correlation has a non zero value at $\delta t=0$. This means that the appearance of a correlation at $\delta t=0$ in the experiment indicates the presence of hydrodynamic interactions.}
	\label{Fig:corr_nume}
\end{figure}

When $k(t)$ is constant with value $k(t) = k_0$, the two-time {cross-correlation} function \begin{equation}
	C_{1,2}(\delta t) = \frac{\langle x_1(0) x_2(\delta t) \rangle}{\sqrt{\langle x_1^2(0) \rangle \langle x_2^2(0) \rangle}}\, 
	\label{eq:corr}
\end{equation}
can be analytically computed from Eq.~\eqref{eq:coupledLangevin} and the results have been experimentally checked \textcolor{black}{\cite{Berut}}, showing excellent  agreement with the theoretical predictions. These equations \eqref{eq:coupledLangevin} can be used to check numerically, in the case of a switching stiffness $k(t)$,   
 how the hydrodynamic coupling modifies the predictions of the theory for totally independent particles obtained in Ref. \cite{BKMS24}. Here we assume that at $\delta t =0$, we start from the stationary state, which could be either equilibrium ($k_1=k_2$) or nonequilibrium ($k_1 \neq k_2$). The effect of the coupling can be tuned in a numerical simulation by changing the value of $\epsilon$. 
In the numerical simulation we use $\gamma=1, k_BT=1$ and $\epsilon=0.17$  which is close to the value between contiguous beads in our experiment. The integration is performed using Euler method with a time step $10^{-2}$.
The results are summarized in Fig.~\ref{Fig:corr_nume} where we plot the two-time cross-correlation function $C_{1,2}(\delta t)$ defined in Eq.~\eqref{eq:corr} as a function of the {time-delay} $\delta t$. There are four curves  that illustrate three different cases. 

\begin{itemize}

\item[(i)] {The flat red curve {corresponds} to the case when $k_1\neq k_2$ and $\epsilon=0$ (no hydrodynamic interaction). In this case, the cross-correlation function in Eq. (\ref{eq:corr}) remains zero at all times, which follows by symmetry, as predicted in our theory \cite{BKMS24}. This is a first check of the theory, which does not have hydrodynamic interaction.} 

\item[(ii)]{The purple curve corresponds to the equilibrium case when $k_1=k_2$ and $\epsilon > 0$ (i.e., in the presence of hydrodynamic interactions). It shows both numerical simulations and the analytical predictions for the equilibrium {time-delayed} correlation function $C_{1,2}(\delta t)$ from \cite{Berut}. In this case, one would expect that, at $\delta t =0$, the cross-correlation should vanish since, at equilibrium, the joint distribution of the positions of the two particles factorizes \cite{Berut}. The fact that the purple curve goes to $0$ as $\delta t \to 0$ confirms this fact. For nonzero $\delta t$, this curve becomes negative and develops a minimum, indicating that an effective attractive correlation develops at early times under the effect of the hydrodynamic interactions. However, from the point of view of this paper, what is really important is the fact that this correlation vanishes at $\delta t =0$. The physics of $\delta t > 0$ is not the focus of the current paper and we refer the interested reader to Ref. \cite{Berut}. 
 }
\item[(iii)] {This case corresponds to the green and the blue curves in Fig. \ref{Fig:corr_nume}. They both correspond to a finite fixed $\epsilon  = 0.17 > 0$ and two sets of stiffnesses $(k_1 = 1.44, k_2 = 0.05)$ (blue) and $(k_1 = 1.44, k_2 = 0.2)$ (green). In this case, the system is in a NESS at $\delta t = 0$ in the presence of hydrodynamic interactions $\epsilon > 0$. In addition, the particles are also correlated in the NESS due to the dynamically emergent correlations (DEC) generated by the switching of the stiffness. Again, focusing on $\delta t = 0$, we clearly see from both curves that $C_{1,2}(\delta t= 0) > 0$. Had $\epsilon$ been equal to $0$, as assumed in the noninteracting theory in this paper, one would have obtained $C_{1,2}(\delta t =0) = 0$, just by symmetry as in the flat red curve. The main conclusion that we should draw from these two curves is that in the NESS and in the presence of hydrodynamic interactions $C_{1,2}(\delta t = 0) >0$, while it would vanish for $\epsilon=0$.}
 \end{itemize}

To summarize, the above exercise with two particles leads us to the following conclusion: the fact that in our experiment with multiple particles we find $C_{1,2}(\delta t = 0)>0$ (see the purple curve in Fig. EM1) clearly indicates that the hydrodynamic interactions are present in the system. Despite the presence of hydrodynamic interactions, the fact that the higher order correlation function (see Fig. 2 in the Letter) as well as the distribution of the extreme $M_1$ in Fig. 3 of the Letter match very well with the theory without hydrodynamic interactions signals that the DEC completely overwhelms the nonzero hydrodynamic interactions. However, quantifying the effects of hydrodynamic contributions to these curves is very hard, as it involves a detailed modelling of the fluid flow around the particles, as well as the boundary conditions, etc. Hence we illustrated this with the help of the example of a two-body problem which brings home this point rather clearly and convincingly.

\begin{figure}[t]
\centering
\includegraphics[width = 0.7 \linewidth]{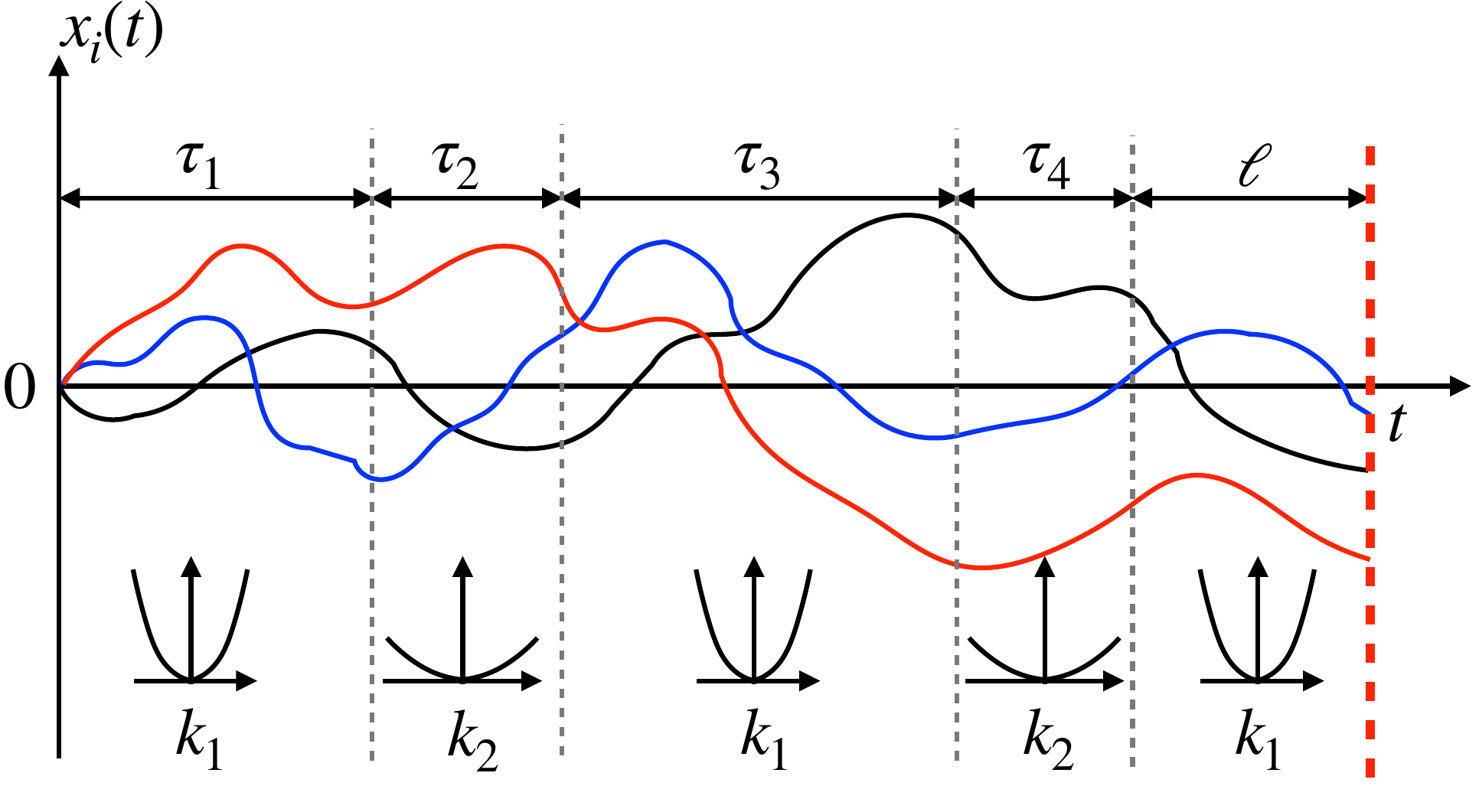}
\caption{Sketch of the switching protocol $\{\tau_1,\tau_2,\dots,\tau_n,\ell,n\}$ with $n=4$, drawn from 
the joint distribution in Eq.~\eqref{eq:prob}. The trajectories $x_1(t), x_2(t), \cdots x_N(t)$ of $N=3$ independent particles are shown in black, blue and red. The switching epochs 
$\tau_1,\ \tau_1+\tau_2,\ \dots,\ \tau_1+\cdots+\tau_n$ mark the instants at which the confining 
potential alternates between $V_1(x)=\tfrac{k_1}{2}x^2$ and $V_2(x)=\tfrac{k_2}{2}x^2$ (or vice versa), 
while $\ell$ denotes the duration of the last incomplete interval preceding time $t$.
} \label{Fig_protocol}
\end{figure}
\section{Switching protocol and the origin of the correlations between particles}
\label{sec:cor}
\textcolor{black}{In the Letter and in the previous sections we studied the system in the steady state. Here we consider a more general situation at a fixed time $t$ where the system has not necessarily reached 
a steady state. This analysis allow us to show  the different  behavior of the correlation  in the case of either stochastic or  periodic forcing  and to quantitatively estimate the contribution of hydrodynamic coupling.}

\subsection{Noninteracting particles}

We start with $N$ noninteracting  particles with coordinates $x_1,x_2\cdots x_N$ evolving stochastically, starting at the origin at $t=0$, via 
\begin{equation}
\label{eq:lan}
    \frac{dx_i}{dt} = -k(t)x_i+ \sqrt{2 k_B T}\,\eta_i(t)\,,
\end{equation}
where $\eta_i(t)$ are independent zero mean Gaussian white noises with correlator $\langle \eta_i(t)\eta_j(t')\rangle = \delta_{i,j} \delta(t-t')$. In Eq.~\eqref{eq:lan}, the stiffness $k(t)$ is a dichotomous telegraphic process switching between two values $k_1$ and $k_2$ (see Fig.~\ref{Fig_protocol}). The stiffness stays at value $k_1$ (or $k_2$) {up to} a random time drawn from $\psi_1(\tau)$ (resp. $\psi_2(\tau)$). Here, for simplicity, we consider the case set $\psi_1(\tau)=\psi_2(\tau)=\psi(\tau)$. However, our analysis below can be easily extended to the more general {asymmetric} case $\psi_1(\tau)\neq\psi_2(\tau)$. 

The Langevin equation governing the motion of each particle changes simultaneously at every switching event. Suppose that in a given interval $(0,t)$ the switching occurs at epochs $\{\tau_1,\tau_1+\tau_2,\cdots \tau_1+\tau_2+\cdots\tau_n\}$ where $n$ denotes the number of switching events (see Fig. \ref{Fig_protocol}). The switching protocol can be characterized by the set of random variables $\{\tau_1,\tau_2\cdots \tau_n,\ell,n\}$ where $\ell=t-\sum_{i=1}^n \tau_i$ denotes the length of the last incomplete interval preceding time $t$. Given the observation time $t$, the joint distribution of the random variables associated with the switching protocol can be written as 
\begin{equation}
\label{eq:prob}
\textrm{Prob.} [\tau_1,\tau_2\cdots\tau_n,n|t] = \psi(\tau_1)\psi(\tau_2)\cdots \psi(\tau_n)\Psi(\ell)\delta\left(\sum_{i=1}^n \tau_i+\ell-t\right)\,\quad  \textrm{where}\,\quad \Psi(\ell) = \int_{\ell}^{\infty} d\tau\, \psi(\tau)\,.
\end{equation}
Here $\psi(\tau)$ denotes the distribution of the interval between two successive switchings. Since the last interval of length $\ell$ is not yet complete it comes with the weight $\Psi(\ell)$. The delta function in Eq.~\eqref{eq:prob} ensure that the total length of the interval is $t$. One can check that the joint distribution in Eq.~\eqref{eq:prob} is appropriately normalized to unity.

There are two sources of randomness in a given trajectory of the process: (a) the stochastic evolution of each trajectory given the switching protocol characterized by the random variables $\{\tau_1,\tau_2\cdots \tau_n,\ell,n\}$ and (b) the random nature of the switching protocol itself. To compute the joint distribution $P(\vec{x},t)$ where $\vec{x}\equiv \{x_1,x_2,\cdots x_n\}$ at time $t$, we proceed in two steps: (1) we first fix the switching protocol $\{\tau_1,\tau_2\cdots \tau_n,\ell,n\}$ and denote the  position distribution conditioned on the protocol by $P_c(\vec{x},t|\tau_1,\tau_2,\tau_3\cdots \tau_n,\ell,n)$ and (2) we average over the protocol variables $\{\tau_1,\tau_2\cdots \tau_n,\ell,n\}$ drawn from the joint distribution in Eq.~\eqref{eq:prob}. We first note that the conditional position distribution $P_c(\vec{x},t|\tau_1,\tau_2,\tau_3\cdots \tau_n,l,n)$ is factorizable at any time $t$ since the particles are independent. This gives 
\begin{equation}
   P_c(\vec{x},t|\tau_1,\tau_2,\tau_3\cdots \tau_n,\ell,n) = \prod_{i=1}^n P_c(x_i,t|\tau_1,\tau_2,\tau_3\cdots \tau_n,\ell,n)\,.
   \label{eq:fac}
\end{equation}

\begin{figure}[t]
    \centering
    \includegraphics[width=0.4\linewidth]{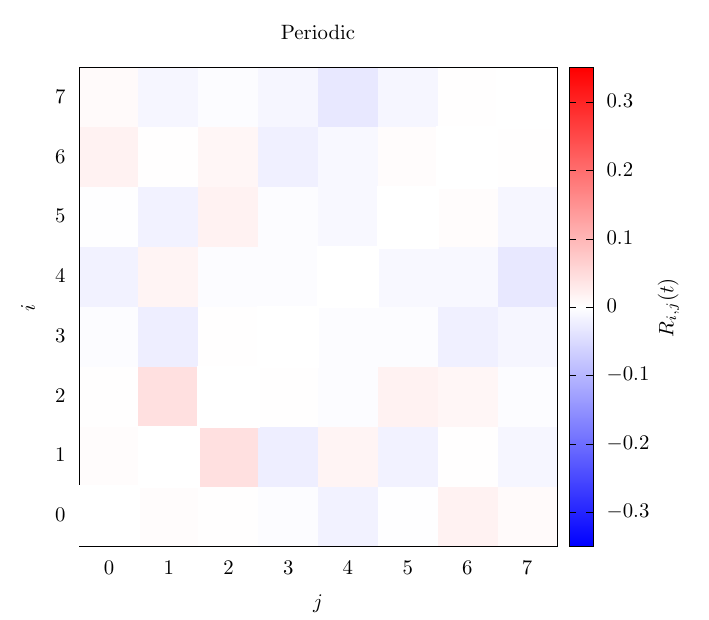}%
    \includegraphics[width=0.4\linewidth]{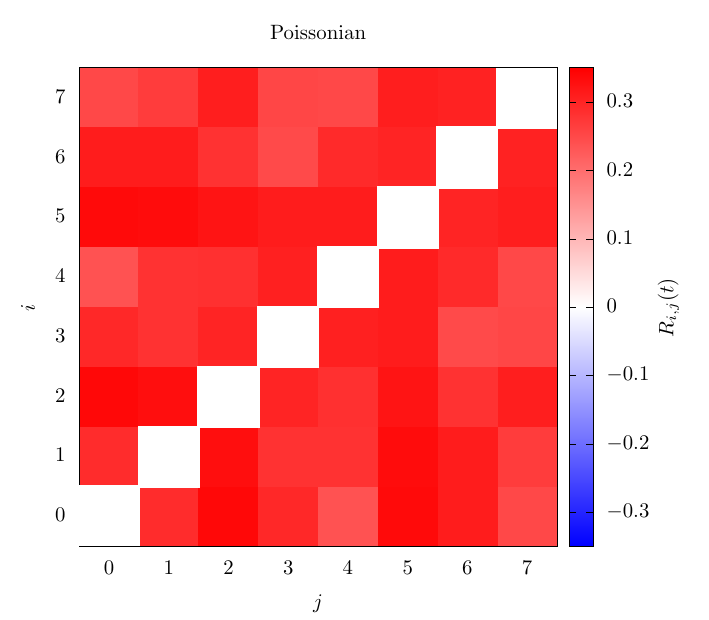}
    \caption{
    \textcolor{black}{Each matrix displays the off-diagonal elements of the normalized correlator matrix $R_{i,j}(t) = \langle x_i^2(t) x_j^2(t) \rangle / (\langle x_i^2(t) \rangle \langle x_j^2(t) \rangle) - 1$, obtained from numerical simulations. On the left, we simulated a periodic switching protocol whereas on the right we simulated a Poissonian switching protocol.
    We obtained $20\,000$ simulated samples for $N = 8$ via Euler-Maruyama integration of $\dd x = -\mu(t) x \dd t + \xi \sqrt{2 D \dd t}$, from $x_0 = 0, t_0 = 0$ to $t_{\rm snapshot} = 20$ with $D = 1.0$ and $\dd t = 10^{-3}$. The potential $\mu(t)$ switches between $\mu_1 = 1.0$ and $\mu_2 = 4.0$ either stroboscopically at time interval $T = 1.0$ or Poissonianly with rate $r = 1.0$.} }
    \label{fig:stroboscopic_vs_poisson}
\end{figure}
Hence the unconditional position distribution can be expressed as 
\begin{equation}
\label{eq:unc}
P(\vec{x},t) = \sum_{n=0}^{\infty} \int_0^{\infty} d\tau_1d\tau_2\cdots d\tau_n\, \textrm{Prob.} [\tau_1,\tau_2\cdots\tau_n,n,\ell|t] \,\prod_{i=1}^n P_c(x_i,t|\tau_1,\tau_2,\tau_3\cdots \tau_n,\ell,n)\,,
\end{equation}
where $\textrm{Prob.} [\tau_1,\tau_2\cdots\tau_n,\ell,n|t]$ is given in Eq.~\eqref{eq:prob}.

\vspace*{0.5cm}
\noindent{\textit{The case of a stroboscopic {switching} protocol.--}} For a generic protocol characterized by $\psi(t)$, it follows from Eq.~\eqref{eq:unc}, that the position distribution $P(\vec{x},t)$ does not factorize at any $t$, indicating that the particles are correlated via the fluctuating switching protocol. However, for the special case of stroboscopic switching with period $T$, there $\psi(\tau)=\delta(\tau-T)$, it follows that for a given $t$ and $T$, the number of resettings $n$ in time $t$ is also unique with a value $n^* = 
{\rm Int}(t/T)$ where ${\rm Int}(x)$ denotes the integer part of $x$. In this case, therefore, there is only a single term in the summation and the integrals in Eq.~\eqref{eq:unc} contributes, leading to 
\begin{equation}
\label{eq:uncst}
P(\vec{x},t) =\prod_{i=1}^n P_c(x_i,t|T,T,T,\cdots T,t-n^{*}T,n^{*})\,.
\end{equation}
Eq.~\eqref{eq:uncst} clearly shows that in the stroboscopic case, the {\color{black} position} distribution remains factorizable at all times $t$. Furthermore, it does not reach a time independent state at long times, but rather a time periodic stationary state (of period $T$). Hence, in this case, the dynamics does not create any correlation between particles. For the particles to be correlated, one needs the switching protocol to be fluctuating, i.e., the interval distribution $\psi(\tau)$ is different from a single delta function as, e.g., in the case of exponential distribution where $\psi(\tau)=r\,e^{-r\tau}$. \textcolor{black}{Thus we conclude that 
for periodic/stroboscopic forcing of noninteracting particles, the joint distribution factorises at all 
times (irrespective of the initial conditions) and not necessarily only at late times.}

\textcolor{black}{The theoretical analysis above clearly shows that, for noninteracting particles, the periodic switching does not generate any correlation, while stochastic switching does produce nonzero correlations. Thus, quite strikingly, one needs a little bit of stochasticity to generate the DEC for noninteracting particles. To test this theoretical prediction, we numerically measured the normalized correlator matrix 
\begin{eqnarray} \label{def_Rij}
R_{i,j}(t)={\langle x_i^2(t) x_j^2(t) \rangle \over \langle x_i^2(t) \rangle \langle x_j^2 (t)\rangle}-1 \quad, \quad 
\end{eqnarray}
both for the periodic as well as the stochastic drive for $N=8$ particles. Note that, for $i \neq j$, all the off-diagonal elements of this matrix are identical and they coincide (in the limit $t \to \infty$) with $C_2$ defined in Eq. (4) of the main text. In Fig. \ref{fig:stroboscopic_vs_poisson} we show a color map of only the off-diagonal elements of this $8 \times 8$ matrix. For the periodic drive, all the off-diagonal elements vanish (shown by the white color), while for the stochastic Poissonian drive, they are identical and nonzero, as shown by the uniform red color.}

\begin{figure}[t]
   \includegraphics[width = \linewidth]{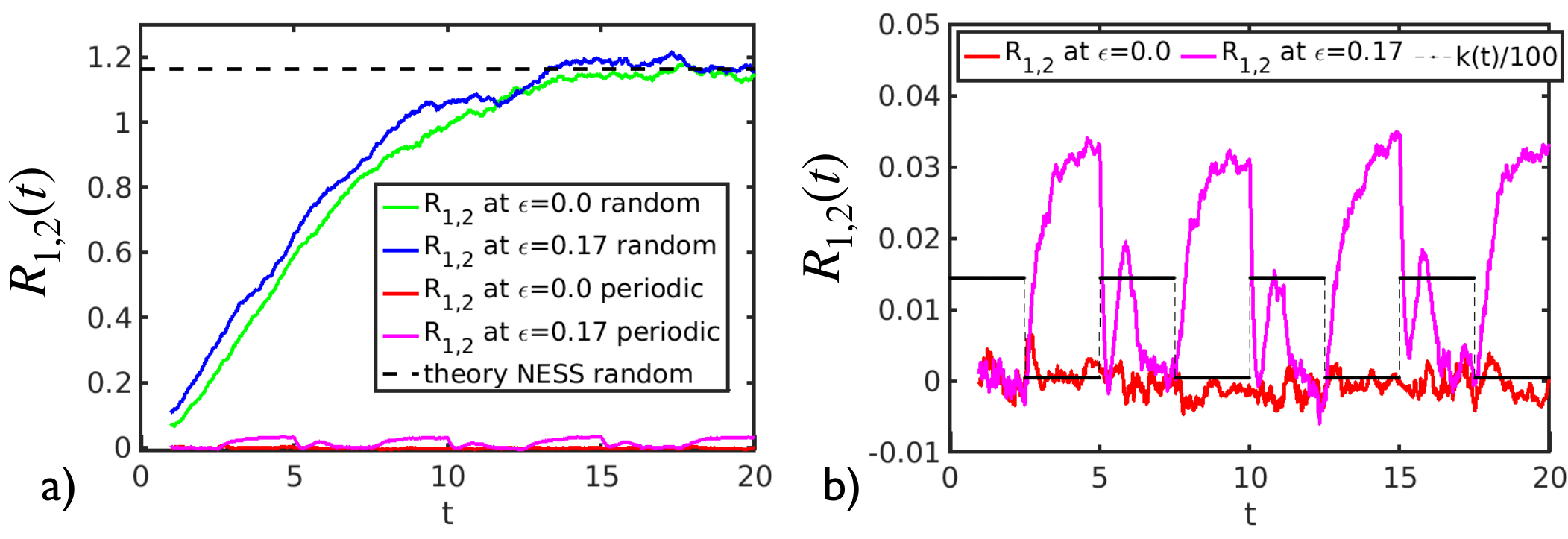} \caption{\textcolor{black}{ a) Plot of $R_{1,2}(t)$ in Eq. (\ref{def_Rij}) vs $t$, resulting from the numerical integration of Eq. \eqref{eq:coupledLangevin}. The curves correspond to the mean of $10^5$ trajectories for Poissonian driving with rate $r=0.2$ and $10^6$ for periodic driving with period $T=5$. The simulations started from the initial condition $x_i(0) = 0$. 
    The horizontal dashed line corresponds to the theoretical prediction for $C_2 = R_{i,j}(t \to \infty)$ given in Eq. (5) of the main text, valid
    in the NESS  of  noninteracting particles driven by a stochastic Poissonian modulation. b) Expanded view of the periodic case in a). The amplitude of the stiffness $k(t)/100$ is also plotted (black line) to point out the synchronization between the driving and the oscillations of  $R_{1,2}(t)$ at $\epsilon\ne 0$. Notice that the effect of the interaction is larger when the stiffness is small. 
    }}
    \label{fig:periodic_random_hydro}
\end{figure}
\textcolor{black}{\subsection{ Interacting particles coupled hydrodynamically}
In order to estimate the contribution of hydrodynamic coupling on $R_{i,j}(t)$ we solve numerically the model described by  Eqs. \eqref{eq:coupledLangevin} which, as we said, has been tested experimentally for the two beads interactions \cite{Berut}. Specifically we start the simulation at $t=0$ and $x_1=0,x_2=0$. We modulate the stiffness from $k_1=1.44$ to $k_2=0.05$ either by Poissonian stochastic switching with rates $r_1=r_2 =r=0.2$, or with a periodic square wave with period $T=5$. We performed simulations with the hydrodynamic coupling $\epsilon$ set to $\epsilon=0$ (noninteracting) and $\epsilon=0.17$ (interacting). This value $\epsilon=0.17$ is close to the one inferred in the experiment (see Sec. \ref{Sec_hyd_coup}). In Fig. \ref{fig:periodic_random_hydro}, we plot $R_{i,j}(t)$ as a function of $t$. Looking at Fig.  \ref{fig:periodic_random_hydro}a), we notice that: (i) 
in the case of stochastic forcing (the blue and the green curves respectively for $\epsilon = 0.17$ and $\epsilon = 0.0$) the hydrodynamic interaction at $\epsilon=0.17$ affects only the growth but the mean steady state value is equal within error bars to the theoretical value for non-interacting beads as in the experiment; (ii) for the periodic drive, the correlations in both the interacting ($\epsilon = 0.17$, purple curve) and the noninteracting case ($\epsilon = 0.0$, red curve) are close to zero.} 

\textcolor{black}{
In Fig. \ref{fig:periodic_random_hydro}b), we zoom out the periodic case and find an interesting structure at a finer scale. The effect of setting  $\epsilon\ne 0$ is clearly important in the periodic forcing plotted, as seen on the purple curve in Fig. \ref{fig:periodic_random_hydro}b). Here we notice that at $\epsilon=0$ $\langle R_{1,2}(t) \rangle \simeq 0\pm 0.002$. Instead the hydrodynamic interactions produce oscillations of $R_{1,2}(t)$ synchronous with the driving. However these oscillations are so small that it is impossible to detect them in the experiment because of statistical accuracy and unavoidable experimental drifts.}

\textcolor{black}{
Let us conclude this subsection with the remark that for 
periodic forcing of interacting particles, there is no NESS at long times in the sense 
that the joint distribution does not approach a time-independent form as $t \to \infty$. However, at late times, it does approach a time-periodic stationary state where the form of 
the joint distribution, though still time-dependent, becomes periodic with time. This is also seen in Fig. \ref{fig:periodic_random_hydro}b).}

\section{Experimental apparatus} 
The experimental set up has been already described in several articles,  e.g., in Refs  \cite{Berut,Baiesi,Berut_2016,Martinez,Berut_2016B,Martinez_2023} where  hydrodynamic interactions of two beads in the presence of external forcing was studied. Thus we report here only the main features. 
\begin{figure}[h!]
	\centering	
	\includegraphics[width=0.6\linewidth]{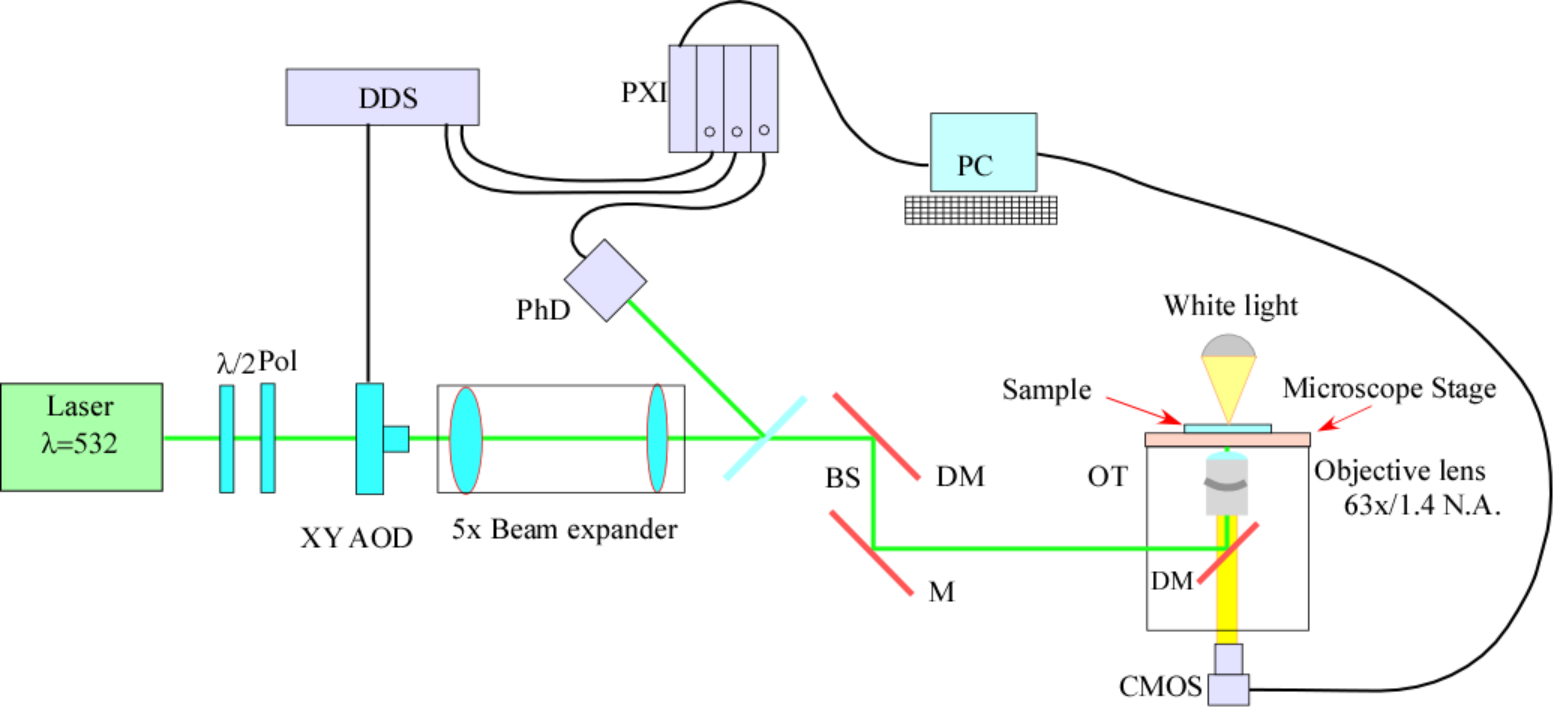}  
	\caption{Experimental set-up consists of a Laser, a $\lambda/2$ plate-polarizer (Pol) set-up,  two AOD for x,y deflection,  a 5X beam expander, a photodiode (PhD),  Dichroic Mirrors, an oil-immersion microscope objective, a diode white light source, a CMOS fast camera, and a microscope stage used to displace the sample in $x,y,z$ directions.} 	 
	\label{Fig:experim} 
\end{figure}
A schematic diagram of the optical set-up used to produce the four modulated optical traps is sketched in Fig.~\ref{Fig:experim}.
The laser is a LASER QUANTUM OPUS 532nm max power 1W. The output power is controlled by the power driver LASER QUANTUM TEM 00, power stability $0.2\%$, noise $0.08\%$, pointing stability $2\mu$rad/°C. The  $\lambda /2$ plate-polarizer set-up is  used  to adjust the light intensity power on the trap.   Acousto-Optic-Modulators (AOD)  are  the model DDSXY-250-532 AA OptoElectronic, driven   by a  DDS variable frequency driver and  controlled by  2 cards NI PXIe-6535. After the AOD the laser beam is expanded 5 times to fill the aperture of the microscope.
An oil imersion objective  (HCX PL. APO $63×/0.6-1.4$) is used to focus the beam  inside a $250\mu$m thick cell which contains the solution of the beads. The cell is shined by a diode light source and the images of the beads are recorded at 1000fps and exposure time 1ms by a CMOS Camera DALSA Falcon 1M120 HG Monochrome (resolution 1024 $\times$ 1024,  pixel pitch 7.4 $\times$ 7.4 $\mu m^2$, 120fps at full resolution). We used a 64 $\times$ 64 px image thus the max fps is about 1500fps. The  trapped particles are  POLYSCIENCES silica beads with a radius $a=(1  \pm 0.05)\mu $m. 
\begin{figure}[h!]
	\centering
	\hspace{-5cm} {\bf a)} \hspace{6cm} {\bf b)}  \\
	\includegraphics[width=0.4\linewidth]{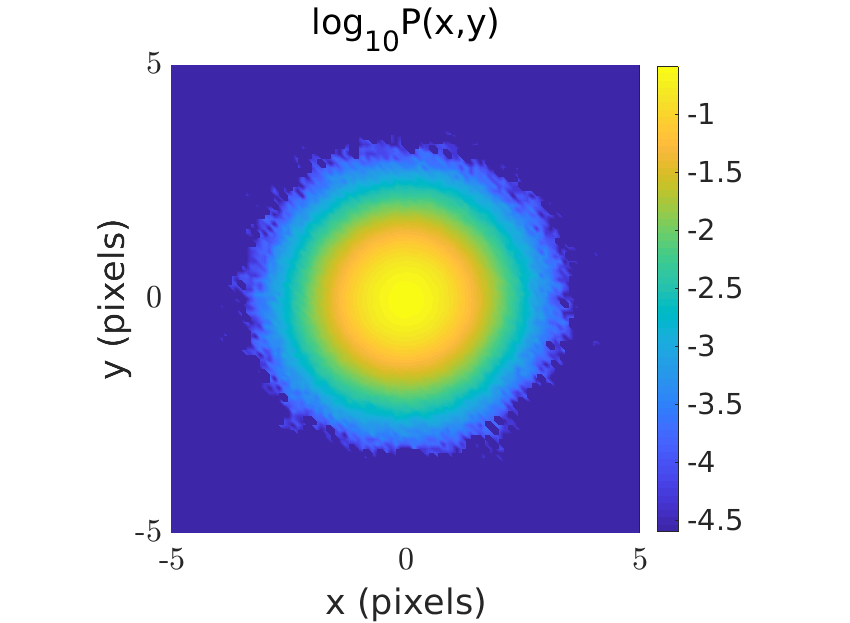} \includegraphics[width=0.4\linewidth]{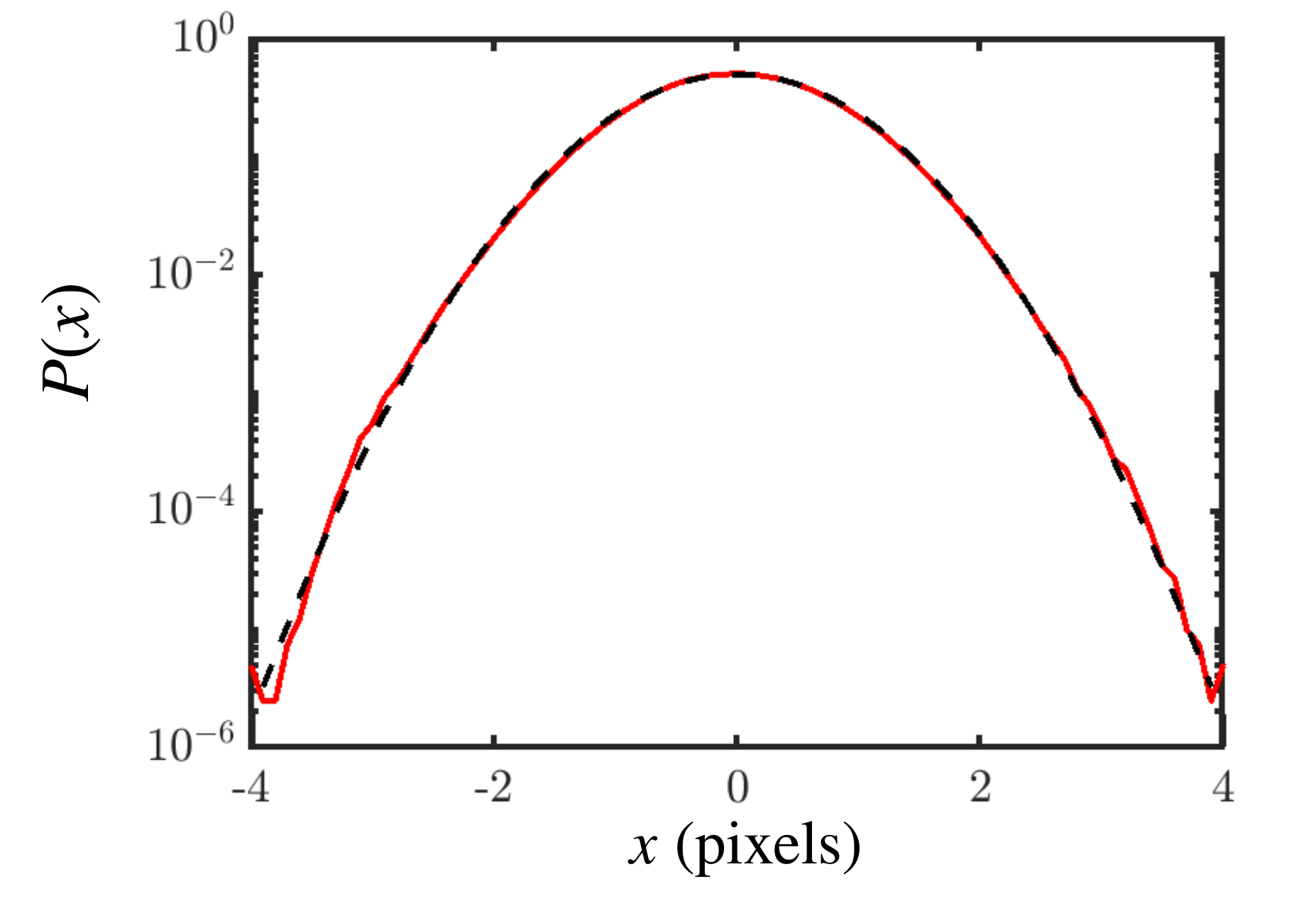} \\
	\hspace{-5cm} {\bf c)} \hspace{6cm} {\bf d)}  \\
	\includegraphics[width=0.4\linewidth]{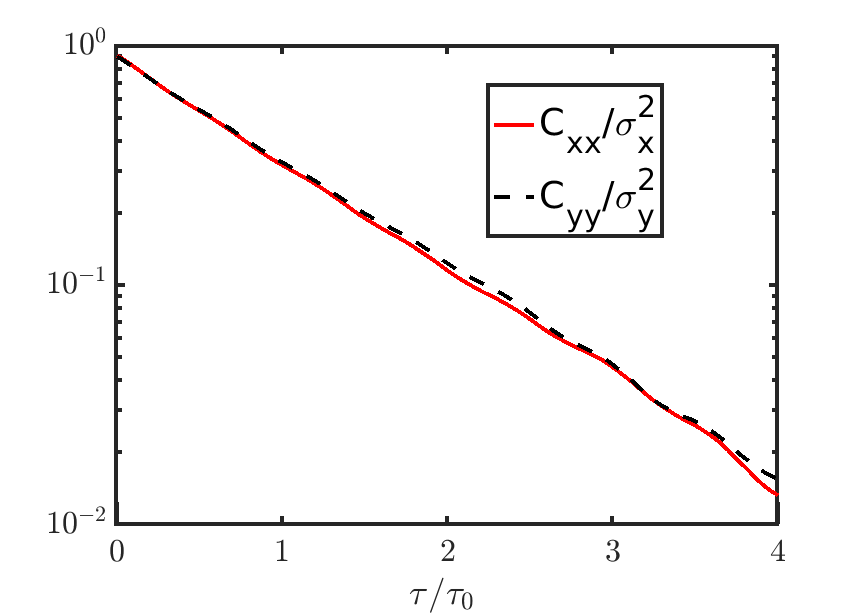} \includegraphics[width=0.4\linewidth]{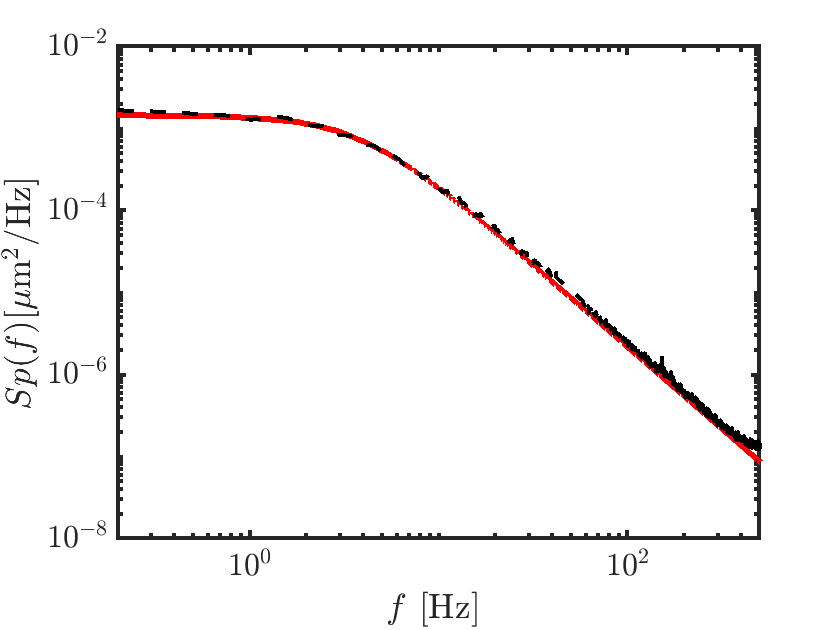}
	\caption{a) The  $P(x,y)$    measured in equilibrium (no forcing) is plotted in log scale. The $P(x,y)$ has been averaged on the 4 beads. This plot shows that the fluctuations in the trap are isotropic and that there is no correlation between x and y. b) $P(x)$ of the 4 particles. The red curve is the experimental pdf and the black line the Gaussian evaluated from the measure of $\sigma$. From the  conversation factor $M_r=0.118\mu m/{pixel}$ ( previously calibrated) we get the calibrated variance and  as a consequence the stiffness $k$. c) The auto correlation functions $C_{xx}$ and $C_{yy}$ are plotted as a function of $\tau/\tau_0$ where $\tau$ is the delay time and $\tau_0=\nu/k$. We use for $k$ the one previously  determined and $\nu = 6\pi \ \eta \ a$. We fit the two correlations function with $\exp(-\alpha \tau/\tau_0)$ finding $\alpha =1 \pm 0.02$. This shows that the calibration of $k$ is correctly done. Of course $\eta$ has been corrected for the temperature and the distance from the bottom plate. d) The estimated value of $k$ and the position calibration has been cross checked by tracing the Lorentian  theoretical spectrum (red curve)
		using the previously estimated values of $k$ and $\nu $. In this figure $k=0.45 \mu$N/m  and $\tau_0=42$ms. 
	} 	 
	\label{Fig:pdf_and_spectre} 
\end{figure}
During sample preparation, the beads are dispersed in bidistilled water at low concentration
to avoid other beads that may be attracted into one of the four confining potentials of our system, thereby perturbing the motion of the already trapped beads. The cell containing the water-beads solution was sealed to avoid evaporation. The top and bottom glass plates of the cell were  respectively $1$mm and  $250\mu m$ thick. For the lateral walls we used  $250\mu$m thick frames (Thermo Scientific™ Gene Frame, $25\mu$l), on which the top an bottom plates are glued  thanks to the the water resistant adhesive deposited on the frames surface (maximum working temperature $90$°C). The distance of the beads from the bottom plate was always larger than $20\mu m$. To further reduce interferences with free diffusing beads during the  measurements, we constructed a small reservoir where the beads are injected into the cell. We took $4$ beads into the reservoir with the optical traps and then we move them by several mm. In such a way we can  perform very long acquisition without any perturbation from other beads which could be attracted into the trap. The experiment was mounted  on a honeycomb optical breadboard with active isolation frame (resonant frequency less than $1$Hz). The table is  in the building basement,in an acoustically isolated and temperature  controlled ($22^o$C) room. 

The particle were tracked in real-time. The tracking program uses  the  centroid based algorithm by  Crocker and Grier described in Ref.~\cite{Grier}. We rewrote  the proposed  matlab  code (https://site.physics.georgetown.edu/matlab/code.html) in C/C++ in order to speed up the acquisition rate.

    The laser power was kept fixed at about 250mW to ensure stability. The intensity modulation was obtained by modulating the driving voltages of the AOD. The light intensity at the exit of the AOD was continuously measured to check the stability of the trapping intensity and the distribution of the driving switching time.  The mean light  intensity on the  focal plane of the microscope was set between 4 and 40mW using the $\lambda/2$-polarizer set up.
	 
	We calibrated the stiffness $k$ with standard techniques \cite{Ashkin_2006,Volpe,Gittes,Capitanio}. To be specific 
    \begin{enumerate}
        \item  $M_r$ is the spatial resolution of the microscope which transform pixel into real lengths. In our case $M_r=0.118\mu$m/pixel
      \item In equilibrium (no external force applied) we checked that the fluctuations of beads were Gaussian. 
      \item  The variance $\sigma$  in nanometers is obtained by $\sigma^2= \sigma_p^2 M_r^2$, where $\sigma_p$ is the standard  deviation measured in unit of pixel of the camera.
      \item The stiffness is : $k=\sigma^2/(k_BT)$.  The value can be checked looking at the spectra
	  and other techniques.
    \end{enumerate}
At the end the stiffness is known with a few percent accuracy (see Refs.~\cite{Berut,Baiesi,Berut_2016,Martinez,Berut_2016B,Martinez_2023}). The relaxation time is measured by fitting the spectrum and the autocorrelation function. (see Fig.~\ref{Fig:pdf_and_spectre}).
	   The spectrum, plotted in Fig.~\ref{Fig:pdf_and_spectre} d), shows that  there are no problems both at low and at high frequencies, the autocorrelation functions have the correct behavior and that the fluctuations in equilibrium are perfectly Gaussian and isotropic [Fig.~\ref{Fig:pdf_and_spectre} a)-b)]. Furthermore we point out that all the curves in the Figs. 2 and 3 of the main text  are not fits but comparisons with theory with no adjustable parameters.

\section{Probability distribution as a function of the number of particles}

\begin{figure}[t]
	\centering
    \hspace{-6cm} {\bf a)} \hspace{7cm} {\bf b)}  \\
{\includegraphics[width=0.4\textwidth, angle=0]{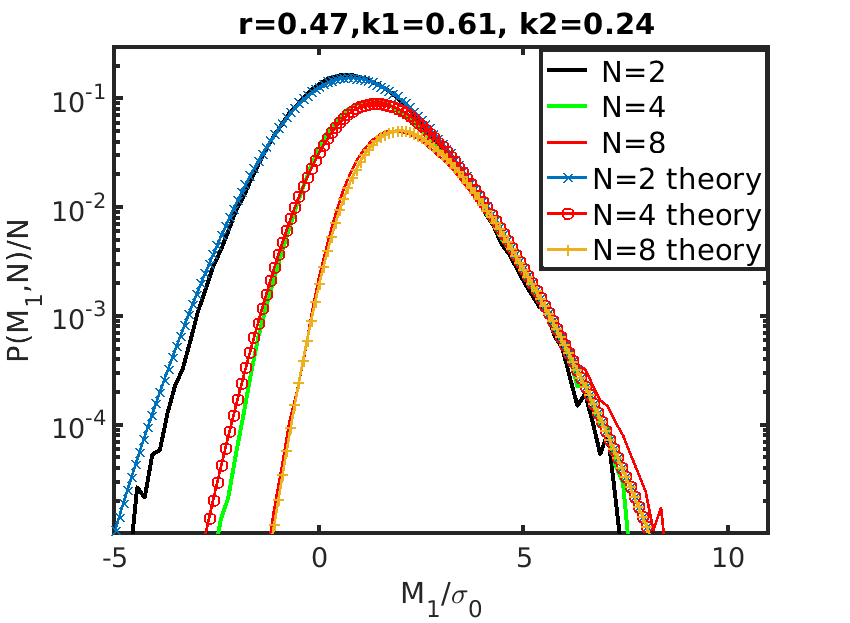} 
		\includegraphics[width=0.4\textwidth, angle=0]{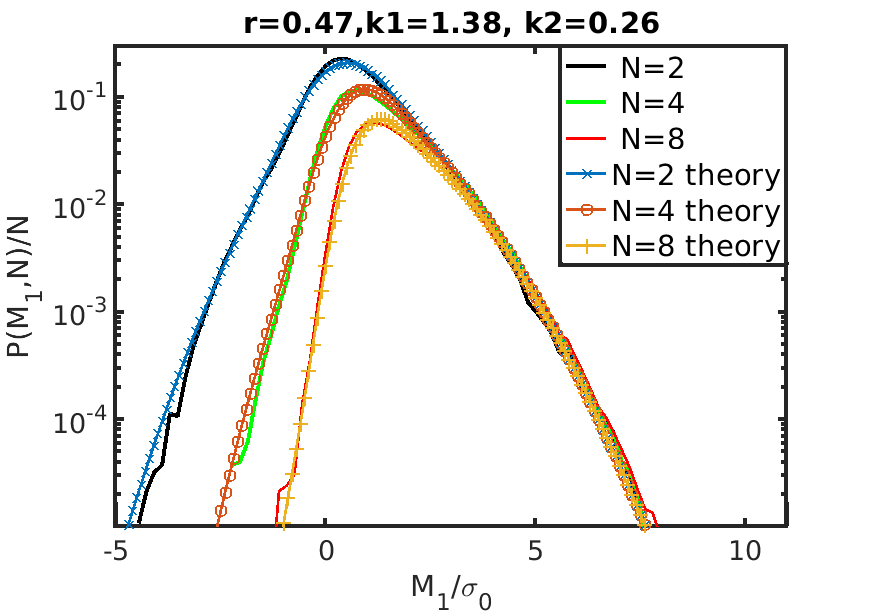} \\
         \hspace{-6cm} {\bf c)} \hspace{7cm} {\bf d)}  \\
		\includegraphics[width=0.4\textwidth, angle=0]{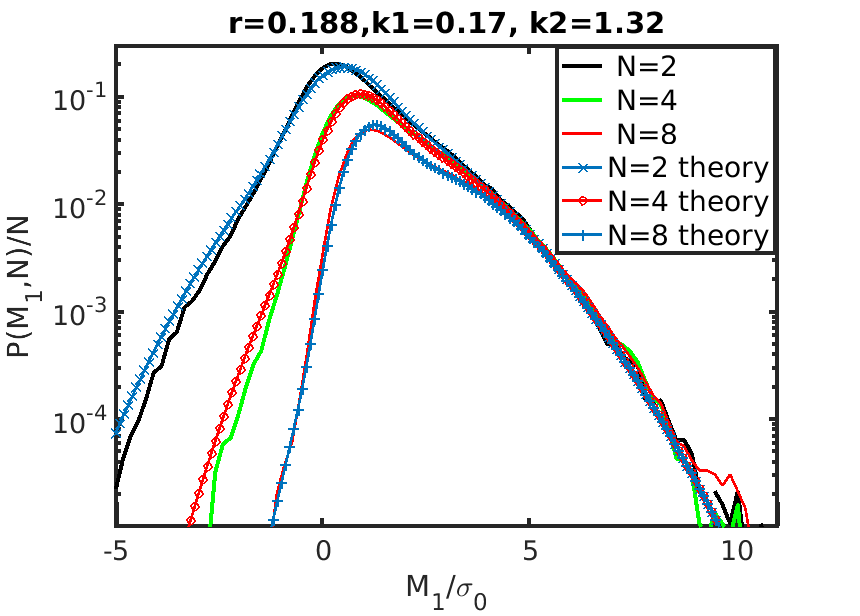}
		\includegraphics[width=0.4\textwidth, angle=0]{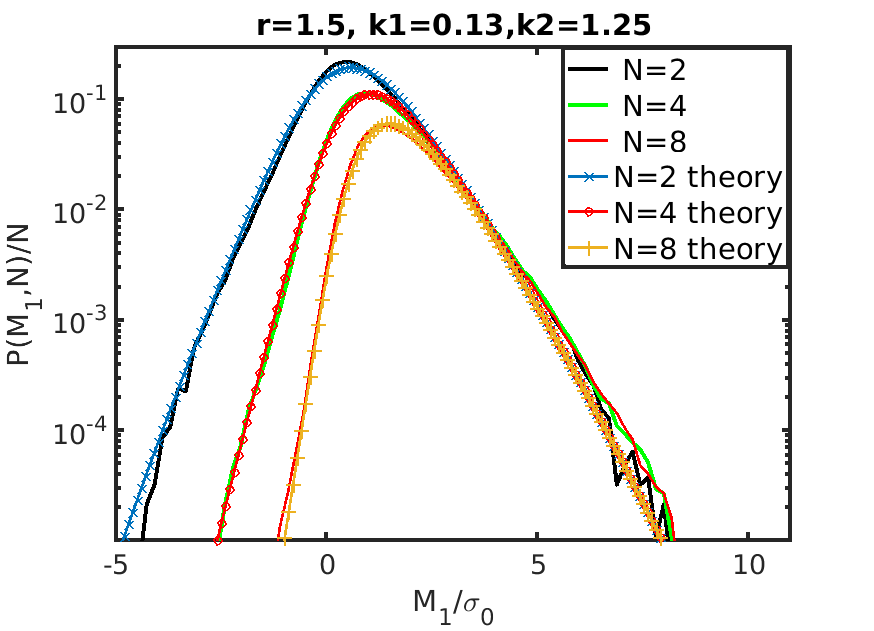}
}
	\caption{The experimental measurements $P(M_1/\sigma_0,N)/N$ for $N=2$,$N=4$ and $N=8$ particles are compared with the theoretical predictions for various $N$.   We see that the agreement with the theoretical predictions is independent of the number of beads $N$. 
	}
	\label{fig:pdf_M1}
\end{figure}
The PDF $P(M_1,N)$ of the position $M_1$ of the rightmost particle has been presented in the main text, where we plot in Fig. 3 the experimental and theoretical results only for 8 beads, i.e., $N=8$. We check here the dependence on $N$ of these PDFs. 
The probabilities $P(M_1/\sigma_0,N)/N$   are plotted in 
Fig. \ref{fig:pdf_M1} for several values of $r,k_1,k_2$  and for $N=2$, $N=4$ and $N=8$ beads. As in the main text the 8 beads are obtained by considering also the $y$ coordinates as they are uncorrelated  with  $x$. We notice that, with increasing $N$, the PDF $P(M_1,N)$ decreases slightly for $M_1<0$. In contrast, for $M_1>0$, the PDF almost remains unchanged. The agreement with the theoretical prediction based on noninteracting model is excellent. Note however that the hydrodynamic interactions are still present, as evident from the measurement of $C_{i,j}(\delta t = 0)$ shown in Figs. \textcolor{black}{EM1 of the End Matter} and \ref{Fig:corr_nume}.    
Thus from these results we conclude that the agreement with the theoretical prediction is excellent, independently of the number of beads and of the values of all  other parameters. 

\vfill
\eject